\documentclass[11pt]{article}
\usepackage{textcomp}
\usepackage{pdfsync}

\usepackage{amssymb}
\usepackage{amsmath}

\usepackage{graphicx}
\usepackage{latexsym}
\usepackage{appendix}
\usepackage{srcltx}
\def\Tr{{\rm Tr}}

\def\CD{{\cal D}}
\def\CE{{\cal E}}

\def\CL{{\cal L}}
\def\CK{{\cal K}}

\def\CS{{\cal S}}
\def\CT{{\cal T}}

\def\centeron#1#2{{\setbox0=\hbox{#1}\setbox1=\hbox{#2}\ifdim
   \wd1>\wd0\kern.48\wd1\kern-.48\wd0\fi
   \copy0\kern-.48\wd0\kern-.48\wd1\copy1\ifdim\wd0>\wd1
   \kern.48\wd0\kern-.48\wd1\fi}}

\def\JHEP{JHEP~}

\def\PRL{Phys. Rev. Lett.~}
\def\PR {Phys. Rev.~}
\def\CQG {Class. Quant. Grav.~}
\def\PL {Phys. Lett.~}

\newcommand{\beq}{\begin{equation}}
\newcommand{\eeq}{\end{equation}}
\newcommand{\bea}{\begin{eqnarray}}
\newcommand{\eea}{\end{eqnarray}}
\newcommand{\ba}{\begin{array}}
\newcommand{\ea}{\end{array}}

\newcommand{\p}{\partial}
\newcommand{\nn}{\nonumber}

\newcommand{\half}{\frac{1}{2}}

\begin{document}

\hskip3cm

 \hskip12cm{CQUeST-2011-0501}
\vskip3cm

\begin{center}
 \LARGE \bf   AdS/BCFT Correspondence for Higher Curvature Gravity: An Example
\end{center}

\vskip2cm

\centerline{\Large \Large Yongjoon
Kwon$^{1}$  \,,~~Soonkeon
Nam$^{2}$\,, ~~Jong-Dae
Park$^{3}$\,, ~~Sang-Heon
Yi$^{4}$ }

\hskip2cm

\begin{quote}
Department of Physics and Research Institute of Basic Science, Kyung
Hee University, Seoul 130-701, Korea$^{1,2,3}$

Center for Quantum Spacetime, Sogang University, Seoul 121-741,
Korea$^4$
\end{quote}

\hskip2cm

\vskip2cm

\centerline{\bf Abstract} We consider the effects of higher curvature terms on a
holographic  dual description of boundary conformal field theory.
Specifically, we consider three-dimensional gravity with a specific
combination of Ricci tensor square and curvature scalar square, so
called, new massive gravity.  We show that a boundary entropy and an
entanglement entropy are given by similar expression with those of
the Einstein gravity case when we introduce an {\it effective}
Newton's constant and an {\it effective} cosmological constant. We
also show that the holographic g-theorem still holds in this
extension, and we give some comments about the central charge
dependence of  boundary entropy in the holographic construction. In
the same way, we consider new type black holes and comment on the
boundary profile. Moreover, we reproduce these results through
auxiliary field formalism in this specific higher curvature
gravity. 
\\
\underline{\hskip12cm}\\
${}^{1}$emwave@khu.ac.kr~~  ${}^{2}$nam@khu.ac.kr~~  ${}^{3}$jdpark@khu.ac.kr ~~ ${}^{4}$shyi@sogang.ac.kr

\thispagestyle{empty}
\renewcommand{\thefootnote}{\arabic{footnote}}
\setcounter{footnote}{0}

\newpage

\section{Introduction}
Anti-de-Sitter/Conformal Field Theory($AdS/CFT$) correspondence has
opened new  and promising  research directions,  after it has been
proposed~\cite{Maldacena:1997re,Witten:1998qj}  as a concrete
realization of holographic
principle~\cite{'tHooft:1993gx,Susskind:1994vu}.  Succinctly
speaking, these directions are related on one side  to studies about
strongly coupled quantum field theories, and, on the other side, to
those about gravity or string theories on AdS space.   While the
original conjecture was about  supersymmetric Yang-Mills theory and
supergravity (or superstrings) on $AdS$ space,  this original form
of the   correspondence has been extended in various ways.  One of
the interesting instances  of $AdS/CFT$  correspondence is the case
of $AdS_3/CFT_2$ correspondence, which is a prime example of how the
black hole entropy may be obtained from field theory
computations~\cite{Strominger:1997eq} and which has been deeply
related  to understanding black hole entropy
microscopically~\cite{Strominger:1996sh}.

Recently, there have been some attempts to extend $AdS/CFT$
correspondence to the case with  boundaries in the CFT
side~\cite{Takayanagi:2011zk}. Specifically, the simplest form of
proposals is that  CFT on a domain $M$ with   conformal  boundary
conditions   on a  boundary $\p M$,  which is called boundary
conformal field theory(BCFT), corresponds to gravity on $AdS$ space
with a  boundary $Q$. The bulk boundary $Q$ is taken on bulk $AdS$
space by demanding  that  its intersection with asymptotic $AdS$
boundary $M$ be given by $\p M$ and it preserves the same amount of
symmetry  with the conformal boundary $\p M$.  To construct  dual of
BCFT, the Neumann boundary conditions are taken on $Q$ with the
introduction of matter stress tensors, while the usual Dirichlet
boundary conditions are done on $M$.

Through this correspondence between two dimensional BCFT and three
dimensional gravity on $AdS$ space with boundary, many interesting
results in $BCFT_2$ are obtained by holographic computations.  The
so-called boundary entropy~\cite{Affleck:1991tk} is   one of the
interesting physical quantities obtained  in this way.  The boundary
entropy  in $BCFT_2$   is defined by logarithm of $g$-function which
is  a disk amplitude of boundary state $|B_\alpha\rangle$ with a
boundary condition $\alpha$. It counts the degeneracy of boundary
ground states and plays a role of   boundary central charge.
Moreover, it satisfies g-theorem~\cite{Friedan:2003yc} which says
that it decreases along renormalization group flow and so is
regarded as the boundary analogue of
c-theorem~\cite{Zamolodchikov:1986gt}. It has been known that the
boundary entropy has interesting connection with the entanglement
entropy which has also a holographic dual description. Consequently,
the boundary entropy was shown to be matched by  independent
computations from holographic dual of disk amplitude and dual of
entanglement entropy.  This agreement gives us some  consistency
checks of various holographic constructions.

On the other hand, there are other interesting directions for the
extension of $AdS_3/CFT_2$ correspondence by considering higher
derivative terms on three-dimensional gravity.  One of them is
specific on three dimensions since it contains three-dimensional
Chern-Simons term~\cite{Deser:1982,Deser:1981wh}. This extension
with some subtleties leads to the chiral gravity conjecture and/or
$\log$-gravity conjecture~\cite{Li:2008dq,Grumiller:2008qz}.  The
others are related to the introduction of higher curvature terms
with specific combination of coefficients which can be extended to
higher dimensions~\cite{Bergshoeff:2009hq}$\sim$\cite{Deser:2011xc}.
Since all these higher derivative gravities allow $AdS$ space as a
solution for some parameter range, one may study the effects of
higher derivative terms on the dual CFT side. One lesson obtained
from these studies on higher curvature theories is that asymptotic
fall-off boundary conditions of bulk fields are important to
characterize theories.

According to these developments in the extensions of $AdS/CFT$
correspondence, one of natural questions  is  what are the effects
of higher curvature terms on the boundary entropy and entanglement
entropy.  As a rule, these  constructions of  BCFT quantities or
$AdS/BCFT$ correspondence  are expected to hold even with higher
curvature theories. Specifically, higher curvature effects on the
entanglement entropy has been studied recently  in several
works~\cite{deBoer:2011wk}$\sim$\cite{Ogawa:2011fw}. In this paper
we focus on three-dimensional gravity with higher curvature terms to
see their effects on the boundary entropy. Our main interest is the
so-called new massive gravity(NMG)~\cite{Bergshoeff:2009hq} \cite{Bergshoeff:2009aq}$\sim$\cite{Nam:2010ma}, which  has a specific
combination of Ricci tensor square and Ricci scalar one.  Since NMG
is  unique in the sense that its Lagrangian is  determined by
holographic c-theorem,  it is a natural playground for $AdS/BCFT$
correspondence.   Interestingly, it turns out that there is an
effective description of NMG which reproduces most of the direct
computations.

This paper is organized as follows. In section 2, we review some
background materials to fix our conventions. We present our
holographic computation of  the boundary entropy in section 3 with
some indications to its extension to even higher curvature terms. In
this section holographic g-theorem for NMG is also derived.  In
section 4, we give some results about black holes and their thermal
properties. We rederive various results by the effective description
of NMG in section 5. In section 6  we summarize our results with
some comments. Various formulae relevant to the main part are
relegated to Appendixes.

\section{A Brief review : boundary entropy in dual BCFT and new massive gravity}
In this section we  review  the holographic construction dual to
BCFT with a boundary $\p M$ following~\cite{Takayanagi:2011zk}  and
also summarize  NMG on three dimensions  to
clarify some points and  to fix our conventions.

\subsection{Holographic boundary entropy}
A boundary entropy is an interesting physical quantity in two
dimensional CFT with boundary, or BCFT (see for a brief
review~\cite{Cardy:2004hm}). It is considered as  the boundary
analogue of  central charge and has an interesting connection with
an entanglement entropy.   To introduce a boundary entropy, let us
consider the disk amplitude of boundary state $|B_\alpha\rangle $ of
a certain BCFT with a boundary condition $\alpha$, which is denoted
as $g_{\alpha}$. Then, the boundary entropy associated with a
boundary is defined by~\cite{Affleck:1991tk}
\beq
 S_{bd} \equiv \ln g_{\alpha} = \ln\, \langle 0 | B_{\alpha} \rangle \,,
\eeq
where $|0\rangle$ denotes the vacuum state. Interestingly, this
boundary entropy is related to the so-called entanglement entropy.

The entanglement entropy of the subsystem $A$ is given by von
Neumann entropy $S_A =  -\Tr \, \rho_A\ln\, \rho_A$ for the reduced
density matrix $\rho_A$ which is the  partial trace of  the total
density matrix $\rho$ with respect to the complement of $A$.   In a
two dimensional CFT defined on a half line  $x <0$, it has been
known that the entanglement entropy of the  finite interval $A: -\mu
\le x <0$  is given by~\cite{Calabrese:2004eu}
\beq
 \CS_A  = \frac{c}{6}\ln \frac{2\mu}{\epsilon} + \ln g + \frac{c'_1}{2}\,, \label{EEA}
\eeq
where $\epsilon$ is UV cut-off, $c$ is the central charge of $CFT_2$, $\ln g$ is boundary
entropy and $c'_1$ is a non-universal constant.  Note that this
result in the two-dimensional CFT is obtained by the so-called
replica method  while the geometric approach to entanglement entropy
(EE) is preferred in generic spacetime dimension. In the geometric
approach it is shown that coefficient of the trace anomaly formula
is related to EE with logarithmic behavior of EE in even dimensions.

Recently, the above two formulae for boundary entropy are realized
holographically as $AdS/BCFT$
correspondence~\cite{Takayanagi:2011zk,Azeyanagi:2007qj}. To
construct these holographic duals to BCFT quantities, it seems
natural to consider Wick-rotated three-dimensional bulk gravity. The
relevant Euclidean bulk action dual to BCFT with one boundary
$\partial M$ for pure Einstein gravity  is given by
\beq S_E = S_{N}  + S_{M} + S_{Q}+S^{mat}_{Q}\,, \eeq
where
\bea S_{N} &=& -\frac{1}{16\pi G} \int_{N} d^3x \sqrt{g} \bigg[ R + \frac{2}{\ell^2} \bigg]\,, \qquad
        S_{M} =  -\frac{1}{8\pi G} \int_{M} d^2x \sqrt{\gamma} K\,,   \label{Eaction} \\
        S_{Q} & =&   -\frac{1}{8\pi G} \int_{Q} d^2x \sqrt{\gamma}  K\,, \qquad  \qquad  \qquad
        S^{mat}_{Q} =   \frac{1}{8\pi G} \int_{Q} d^2x \sqrt{\gamma}~ T\,.  \nn   \eea
In the recent proposal on $AdS/BCFT$ correspondence, the boundary
conditions on $AdS$ boundary $M$ are taken as the usual Dirichlet
ones while those on bulk boundary $Q$ which corresponds to the CFT
boundary $\p M$ are done as the  Neumann ones.  To be a holographic
dual of BCFT, the boundary $Q$ in the bulk $AdS$ space has isometry
dual to conformal symmetry preserved by the CFT boundary conditions.
Note that there are two terms on the boundary $Q$, one of which is
the usual Gibbons-Hawking(GH) term and the other represents the
matter contribution localized on $Q$. For a matter contribution
written by $T$ the boundary condition on the bulk boundary $Q$ is
taken by
\beq 8\pi G T^{ij}_{Q} =  T\gamma^{ij}\,, \qquad
 8\pi G\, T^{ij}_{Q} \equiv  K\gamma^{ij} - K^{ij}\,, \label{bcQ}
\eeq
where $T^{ij}_{Q}$ is the so-called Brown-York stress
tensor~\cite{Brown:1992br,Balasubramanian:1999re} on $Q$ and $K_{ij}$ denotes extrinsic
curvature for $Q$ (see below for our convention). Though the scalar
function $T$ does not need to be a constant in general, the
conservation of Brown-York stress tensor requires the constancy of
$T$. We will give some comments about this requirement in the next
section.

As the simplest construction, one may try to preserve $SO(2,1)$
symmetry  by the boundary conditions among  the full $SO(2,2)$
symmetry of bulk  $AdS_3$ space.  As holographic construction this
symmetry needs to be preserved by the bulk boundary $Q$ as
isometries. Explicitly, $AdS_3$ space in Poincare coordinates $(z,
\tau, x)$ can be represented  as $AdS_2$ fibration over a line    in
coordinates $(\rho, \tau, y)$  as
\beq\label{metric}
 ds^2 = \frac{L^2}{z^2}\Big[dz^2+d\tau^2 + dx^2\Big] =  d\rho^2
     + \cosh^2\Big(\frac{\rho}{L}\Big) \bigg[\frac{L^2}{y^2}(d\tau^2 + dy^2)\bigg] \,,
\eeq
where two coordinates are related by coordinate transformations
\[ z \equiv \frac{y}{\cosh(\rho/L)}\,, \qquad x\equiv y\tanh\frac{\rho}{L}\,, \qquad y \ge 0\,.
\]
Then, the relevant boundary $Q$ can be taken as the $AdS_2$
hypersurface given by $\rho=\rho_*$  with an induced metric $ds^2 =
(L^2/y^2)(dy^2 + d\tau^2)$ and the bulk domain $N$ is specified by
the range $ -\infty < \rho < \rho_*$.  This bulk boundary $Q$ can
also  be represented by $(\tau,z,x(z))$ in original Poincare
coordinates  with
\beq
 x(z) = z\sinh\Big(\frac{\rho_*}{L}\Big) = z\, \frac{TL }{\sqrt{1-T^2L^2}}\,, \label{profx}
\eeq
where we have used  the boundary condition on $Q$ given  in
eq.~(\ref{bcQ}) to replace $\rho_*$ by $T$  in the second equality.
One can see that the domain of BCFT is given by $x < 0$ from this
equation, since   the asymptotic boundary  $M$  is specified by
$z=0$ while the intersection of $Q$ and $M$ is given by $x(0)=0$ and
$(z, x) \rightarrow (0,  -y_0)$ for $(\rho, y)\rightarrow (-\infty,
y_0)$.  As usual, we take $z=\epsilon$ as the geometric cut-off
corresponding to field theory UV cut-off.

After an appropriate conformal transformation from the half space $x
<0$ to a round disk  $\tau^2 + x^2 \le r^2_D$, it turns out that the
holographic dual of this  round disk  in the BCFT  is given by the
domain $N$ in the bulk $AdS_3$ as
\beq
 \tau^2 + x^2 + \Big(z-r_D\sinh\frac{\rho_*}{L} \Big)^2 -r^2_D\cosh^2\frac{\rho_*}{L} \le 0\,,
\eeq
which is a suitable form for obtaining boundary entropy holographically as a disk amplitude.

\subsection{New Massive Gravity}
New massive gravity(NMG) is a three-dimensional gravity with higher
curvature terms, which is regarded as the covariant completion of
Pauli-Fierz massive graviton theory~\cite{Bergshoeff:2009hq}.
Furthermore, it is shown that NMG is the unique extension of
Einstein gravity consistent with holographic
c-theorem~\cite{Sinha:2010,Myers:2010tj}. The Lagrangian of NMG we
will consider in the following is given by
\begin{equation}\label{NMG}
S =\frac{1}{16\pi G}\int d^3x\sqrt{-g}\bigg[ \sigma R +
\frac{2}{\ell^2} + \frac{1}{m^2}\CK  \bigg]\,,
\end{equation}
where   $\sigma$ takes $1$ or $-1$    and $\CK$ is a specific
combination of scalar curvature square  and Ricci tensor square
defined by \beq
 \CK = R_{\mu\nu}R^{\mu\nu} -\frac{3}{8}R^2\,.
\eeq
Our convention is such that the parameter $m^2$  and
cosmological constant $\ell^2$ can take positive or negative values. The equations
of motion  of NMG are given by
\begin{equation}\label{eom}
\CE_{\mu\nu} \equiv
 \sigma G_{\mu\nu} - \frac{1}{\ell^2}g_{\mu\nu} + \frac{1}{2m^2}\CK_{\mu\nu}
        =0\,,
\end{equation}
where ${\cal K}_{\mu\nu}$ is
\begin{equation}
 \CK_{\mu\nu} = g_{\mu\nu}\Big(3R_{\alpha\beta}R^{\alpha\beta}-\frac{13}{8}R^2\Big)
                + \frac{9}{2}RR_{\mu\nu} -8R_{\mu\alpha}R^{\alpha}_{\nu}
                + \half\Big(4\CD^2R_{\mu\nu}-\CD_{\mu}\CD_{\nu}R
                -g_{\mu\nu}\CD^2R\Big)\,. \label{Ktensor}
\end{equation}
Here,  $\CD_{\mu}$ denotes   a  covariant derivative  with respect to $g_{\mu\nu}$.

Now,  let us  consider  the generalized Gibbons-Hawking(GH) boundary
term in NMG. To obtain the generalized GH term in NMG,  it is useful
to introduce an  auxiliary field.    In summary,   the above NMG
action can be rewritten in terms of auxiliary field $f_{\mu\nu}$ as
\beq S =\frac{1}{16\pi G}\int d^3x\sqrt{-g}\bigg[ \sigma R +
\frac{2}{\ell^2} +
f^{\mu\nu}G_{\mu\nu}-\frac{m^2}{4}\Big(f^{\mu\nu}f_{\mu\nu}
    - f^2\Big)  \bigg]\,.  \label{Action} \eeq
In this representation the EOMs of this action are given by
\beq    \sigma G_{\mu\nu} -\frac{1}{\ell^2}g_{\mu\nu}
             =   \CT^{B}_{\mu\nu}\,, \qquad   f_{\mu\nu}
             =\frac{2}{m^2}\Big(R_{\mu\nu} - \frac{1}{4}R g_{\mu\nu}\Big)\,,
        \eeq
where
\bea \!
     \CT^{B}_{\mu\nu} &=& \frac{m^2}{2}\Big[ f_{\mu\alpha} f^{\alpha}_{\nu} -f f_{\mu\nu} -\frac{1}{4}\Big(f^{\alpha\beta}f_{\alpha\beta} -f^2\Big)  g_{\mu\nu}\Big]  + \half f R_{\mu\nu} - \half Rf_{\mu\nu} -2 f_{\alpha (\mu}G^{\alpha}_{\nu)} + \half f^{\alpha\beta}G_{\alpha\beta} g_{\mu\nu} \nn \\
      && -\half \Big[ \CD^2f_{\mu\nu} + \CD_{\mu}\CD_{\nu} f -2\CD^{\alpha}\CD_{(\mu}f_{\nu)\alpha} + \Big(\CD_{\alpha}\CD_{\beta}f^{\alpha\beta} - \CD^2f\Big)g_{\mu\nu} \Big]\,.  \nn \eea

This form of NMG is useful to obtain the generalized GH  term.
To specify various quantities in GH term, we take the relevant metric in the ADM-decomposed form as
\beq ds^2 = N^2d\eta^2 + \gamma_{ij}(dx^i + N^id\eta)(dx^j + N^j d\eta)\,.  \label{ADMdec}\eeq
According to this metric decomposition, the auxiliary field $f^{\mu\nu}$ can be decomposed as
\bea\label{auxliary} f^{\mu\nu} = \left(\ba{cc} s & h^j \\ h^i
& f^{ij} \ea\right) \,.  \nn \eea
With this decomposition the generalized GH term was obtained   in the form of
\beq \label{GH}
 S_{GH} = \frac{1}{16\pi G}\int d^2x \sqrt{-\gamma}\Big[2\sigma K + \hat{f}^{ij}K_{ij} - \hat{f}K\Big]\,,
\eeq
where  $\hat{f}^{ij}$ and $\hat{f}$ are  defined by
\cite{Hohm:2010jc}
%
%
\[
 \hat{f}^{ij} \equiv f^{ij} + 2 h^{(i}N^{j)} + sN^iN^j\,, \qquad  \hat{f}\equiv\gamma_{ij}\hat{f}^{ij} \,.
\]
Note that the first term proportional to $\sigma$ is the GH term in
pure Einstein gravity case and  our convention for the extrinsic
curvature is
\beq \label{ext-def}
 K_{ij} = \frac{1}{2N}\Big( \p_{\eta}\gamma_{ij} - \nabla_iN_j - \nabla_jN_i\Big)\,,
\eeq
where $\nabla_i$ denotes a covariant derivative for the metric $\gamma_{ij}$.

\section{Boundary entropy and holographic g-theorem}

\subsection{Boundary Entropy as a disk amplitude}
As in  the Einstein gravity case we consider $AdS$ space with its
asymptotic boundary $M$ and bulk boundary $Q$   to realize boundary
entropy of BCFT dual to higher curvature gravity. By omitting the
$AdS$ boundary $M$ which leads to counter terms for the divergence cancelation, one may  denote the Euclidean action as
\beq
 S_{E} = S + S_{Q}+S^{mat}_{Q}\,, \eeq
where
\bea
   S    &=&  -\frac{1}{16\pi G} \int_N \sqrt{g}
      \bigg[ \sigma R + \frac{2}{\ell^2} + f^{\mu\nu}G_{\mu\nu}-\frac{m^2}{4}
      \Big(f^{\mu\nu}f_{\mu\nu} - f^2\Big) \bigg]\,,  \label{action-nmg-1}   \\
    S_Q  &=& -\frac{1}{16\pi G} \int_Q \sqrt{\gamma}
      \Big[2\sigma K + \hat{f}^{ij}K_{ij} - \hat{f}K \Big]\,, \qquad
     S^{mat}_{Q} =  \frac{1}{8\pi G} \int_Q \sqrt{\gamma}~ T\,, \label{action-nmg-2}
\eea
and $T$ denotes the contribution from  some matter fields  localized on the bulk boundary
surface $Q$. Though the function $T$ denotes just the Lagrangian for matters on $Q$ in the generic case, we confine ourselves to a simple case such as the scalar function $T$ without metric dependence, which might be realized as `infinitely massive' scalar fields with potential\footnote{The kinetic term of scalar fields are effectively dropped by `infinte mass'.  One may  also recall  that the nature of   higher derivative terms might be different between  gravity and matters in three dimensions. As was realized in three-dimensional TMG or NMG, higher derivative terms in these gravity theories can be made harmless and  have the same weight as the Einstein-Hilbert term, which might be problematic in matter fields  by the existence of ghosts.    This constrasts with the standard string theory approach, in which  all higher derivative terms are treated as small corrections and then those in gravity and matters can be treated on the same footing.}. This means that a certain matter might be  distributed on the surface $Q$, which is specified by the consistency with equations of motion or boundary condition on $Q$. The simplest case is given by the constant matter density $T$, which corresponds to the boundary cosmological constant  or constant scalar potential localized in $Q$ and is the our main focus.

The Brown-York stress tensor for the above action with the convention $T_{ij} =
-\frac{2}{\sqrt{\gamma}} \frac{\delta S}{\delta \gamma^{ij}}$  in
Euclidean signature  is  given by
\beq\label{B-condi}
  8\pi G\,  T^{ij}_{Q} = \Big( \sigma + \frac{1}{2}\hat{s} - \frac{1}{2}\hat{f} \Big)
    (K\gamma^{ij}-K^{ij}) - \nabla^{\left(i\right.} \hat{h}^{\left. j\right)}
    + \frac{1}{2}D_\eta \hat{f}^{ij} + K^{\left(i\right.}_k
    \hat{f}^{\left. j\right) k} + \gamma^{ij} \Big( \nabla_k  \hat{h}^k
    - \frac{1}{2} D_\eta\hat{f} \Big)\,,   \eeq
where hatted quantities are defined by
\beq  \hat{s}=N^2 s\,, \quad   \hat{h}^i = N (h^i+sN^i)\,,  \quad
\hat{f}^{ij} = f^{ij} + 2 h^{(i}N^{j)} + sN^iN^j\,, \quad  \hat{f} =
\gamma_{ij}\hat{f}^{ij}\,, \label{hatfuc} \eeq and the covariant
derivatives along $\eta$ are
\bea
D_\eta \hat{f}^{ij} &=& \frac{1}{N}\Big(\partial_\eta\hat{f}^{ij} -N^k\p_k\hat{f}^{ij} + \hat{f}^{kj}\p_kN^i + \hat{f}^{ik}\p_kN^j\Big)\,, \nn \\
D_{\eta}\hat{f} &=& \frac{1}{N}\Big(\p_{\eta}\hat{f} -N^j\p_j\hat{f}\Big)\,.   \eea
By imposing the Neumann boundary condition on the bulk boundary $Q$,
one obtains the boundary condition on the surface $Q$ as
\beq 8\pi G T^{ij}_Q = T\gamma^{ij}\,.   \label{boundary} \eeq
Noting a useful relation
\[
 D_{\eta} \hat{f} = \gamma_{ij}D_{\eta}\hat{f}^{ij}  + 2
 K_{ij}\hat{f}^{ij}\,,
\]
one can see that the tension of boundary $Q$ is given by
\beq
   2 T  = 8\pi G T^{ij}_Q \gamma_{ij}=\Big( \sigma + \frac{1}{2}\hat{s} - \frac{1}{2}\hat{f} \Big)\, K   - \frac{1}{2}D_{\eta} \hat{f} \,.
\eeq
It is also interesting to see that the Brown-York stress tensor of
NMG already indicates the possibility of its non-conservation
because of various hatted quantities enters in its expression. In
the case of Einstein gravity the so-called second  Gauss-Codacci
equation implies the conservation of Brown-York tensor in some
cases, for instance BTZ black holes~\cite{Banados:1992wn}.
Explicitly, the second Gauss-Codacci equation     is  given    in
our convention   by
\beq\label{Gauss-Codazzi}
   \nabla^j(K\gamma_{ij} - K_{ij}) = -NR^{\eta}_{\, i}\,,  \eeq
where $R^{\eta}_{i}$ is $\eta i$ component of the bulk Ricci tensor.
The conservation of Brown-York tensor charges at the asymptotic infinity  is a usual assumption  in $AdS/CFT$
correspondence, which is related to the conformal invariance in the
CFT side.    The conservation of 
Brown-York tensor  charges at the bulk boundary $Q$ might also  be related to conformal symmetry, which    implies the constancy
of the tension of $Q$ by the boundary condition on $Q$.      Even for BTZ black holes in NMG, the
conservation of Brown-York tensor charges  can be seen  in the same way with
Einstein case in spite of the hatted quantities. However, that is not the case for the 
so-called new type black holes which  are allowed  at special value of
parameters in NMG.   As was
shown in~\cite{Kwon:2011jz,Kwon:2011ey} these black holes  are stable  and
compatible with holographic renormalization, but they are
mysterious in the view point of the AdS/CFT correspondence and  their relevance in
dual CFT is still unclear as was point out in~\cite{Kwon:2011ey}.
Therefore, we study the possibility of the AdS/BCFT correspondence
without the conservation of Brown-York tensor charges keeping
these new type black holes in mind.

Since all the formulae in NMG as well as GH terms and the above
tension $T$ are written in terms of the ADM-decomposed metric, it is
useful to rewrite $AdS$ metric in the ADM-decomposed form.
Specifically, the $AdS_3$ metric in coordinates $(\rho, \tau, y)$ in
eq.~(\ref{metric}) is already in the ADM-decomposed form.

Explicitly,
the on-shell value of auxiliary field
$f_{\mu\nu}$ on $AdS$ space is given by
\[
 f_{\mu\nu} =   -\frac{1}{m^2L^2}\bar{g}_{\mu\nu}\,,
\]
where $\bar{g}_{\mu\nu}$ denotes the background $AdS$ metric. Then,
the hatted quantities for the above ADM-decomposed metric are given
by
\beq \hat{s} = -\frac{1}{m^2L^2}\,, \quad \hat{h}^i =0\,, \quad \hat{f}^{ij} = -\frac{1}{m^2L^2}\gamma^{ij}\,, \quad \hat{f} = -\frac{2}{m^2L^2}\,. \label{auxvalue}\eeq
One may note that $L$ is not identical with the cosmological constant $\ell$ in NMG.

Let us take the same boundary $A: \rho =\rho_*$ with the Einstein
case for the computation of boundary entropy in the holographic
setup. By a straightforward computation, one obtains the total
action value or the disk amplitude in NMG with a boundary $A:
\rho=\rho_*$ as
\beq
 S_{E} = \frac{L}{4G} \Big( \sigma + \frac{1}{2m^2L^2} \Big)
   \bigg[ \frac{r_D^2}{2\epsilon^2} + \frac{r_D \sinh\frac{\rho_*}{L}}{\epsilon}
   + \log \Big( \frac{\epsilon}{r_D} \Big) - \frac{1}{2} - \frac{\rho_*}{L}
   \bigg] \,.  \label{diskamp}
\eeq
Now, one can see that the (regularized) boundary entropy for the boundary $A$  is given by
\beq {\cal S}_{bd} =  S_E(0)- S_E(\rho_*) = \frac{\rho_*}{4G} \Big(
\sigma + \frac{1}{2m^2L^2} \Big)
     = \frac{\rho_*}{4G_{eff}}\,, \qquad G_{eff} \equiv \frac{1}{\big(\sigma + \frac{1}{2m^2L^2}\big)} G\,.  \label{bdent}
\eeq
Since the central  charge of CFT dual to NMG is given by
\footnote{See,  for example
\cite{Bergshoeff:2009aq,Nojiri:1999mh,Liu:2009bk} for this
expression of central charge of dual CFT for NMG.}
\beq c = \frac{3L}{2G}\Big(\sigma + \frac{1}{2m^2L^2}\Big)\,, \label{NMGcen} \eeq
this boundary entropy can be written as \footnote{One may note that
$\rho_*$ has dimension of length.}
\beq  {\cal S}_{bd} = \frac{c}{6}\frac{\, \rho_{*}}{L}\,, \eeq
which is the same form with Einstein gravity except the adaptation
of the central charge to NMG. The behavior of boundary entropy
proportional to central charge is expected to be a universal one in
the holographic construction of boundary entropy.
(see~\cite{Chiodaroli:2011fn} for another example in six dimensional supergravity
setup.)

\subsection{Boundary entropy from holographic entanglement entropy}
The holographic realization of the entanglement entropy  suggests
that higher curvature terms would require some modifications of the
original proposal  on holographic entanglement entropy(HEE) in the
pure Einstein gravity case. Recalling that the original proposal for
HEE is  given by the minimized area of relevant (hyper)surface $S_A$
in the  Euclideanized bulk space, one  naturally expects the
minimization of  some functional defined on $S_A$ would be a correct
way to make modifications. There are some suggestions to identify
this functional at least for the so-called Lovelock
gravity~\cite{deBoer:2011wk,Hung:2011xb}.  One of the results
in these studies is that the Wald formula type of entropy is not a
correct functional for generic higher curvature  gravity.

One may note that  the Wald type of entropy is shown to be
inadequate because of the discrepancy of  HEE and Wald entropy  in
the dependence of  various  central charges  in the trace anomaly.
However, in the $AdS_3/CFT_2$ case of our interest, there is just
single central charge $c$, which  enters at the trace anomaly
formula as
\[ \langle T^{i}_{~\, i} \rangle = \frac{c}{24\pi }R\,. \]
Therefore, we propose  the Wald type of entropy as a correct
HEE functional in our case. More concretely, we propose that Wald
entropy is the correct way to compute HEE for any configuration in
NMG with Brown-Heannaux fall-off boundary conditions. The necessity
of these boundary conditions and the generality of proposal will be
explained in section 5.  In this section, we adopt this prescription
for HEE and test its validity for a specific configuration. A
convenient tool for the computation of  Wald type of entropy  is a
specific central charge $a^*_d$  introduced
in~\cite{Myers:2010tj,Myers:2010xs} as \beq a^*_d =
\frac{\pi^{d/2}L^{d+1}}{d~\Gamma(d/2)}\CL\Big|_{AdS}\,. \eeq
One can see that $a^*_2 =  c/12$ in the case of $d=2$. Then, Wald
type of entropy as HEE is already computed in~\cite{Myers:2010tj}
and reduces in our case to
\beq \CS_{HEE} = S_W = \frac{2}{L}\, a^*_2 \int_{S_A} dx \sqrt{\gamma}   = \frac{c}{6L}\int_{S_A} dx \sqrt{\gamma}\,. \eeq

Now, let us use this formula to compute HEE in our setup and verify
that the relation between HEE and  boundary entropy is not modified
even with higher curvature terms. One may regard this
computation as the verification of our prescription for HEE in our
configuration, since in three dimensions the relation between
entanglement entropy  and boundary entropy is completely determined
by conformal symmetry up to some constant.  For the zero time slice
$\tau=0$, the bulk integration region is given by the same way with
the Einstein case as $-\infty < \rho \le \rho_*$. Therefore, with
some cutoff $\rho_\infty=L\ln (2\mu/\epsilon)$, one obtains
\[ \CS_{HEE} = \frac{c}{6L}\int^{\rho_*}_{-\infty} d\rho = \frac{c}{6L}(\rho_* + \rho_{\infty})\,, \]
which  is  the holographic construction for EE in Eq.~(\ref{EEA})
and leads to the boundary entropy as
\beq {\cal S}_{bd}  = \CS_{HEE}(\rho_*) - \CS_{HEE}(0) =
\frac{c}{6}\, \frac{\,\rho_*}{L} = \frac{\rho_*}{4G'}\,. \eeq
One can see that the complete match of this result  with
Eq.(\ref{bdent}) which is obtained in a different way.

\subsection{Holographic g-theorem for higher curvature gravities}
One interesting quantity in BCFT is the so-called  $g$-function
which is a monotonic function characterizing  the boundary
renormalization group flow and coincides with the boundary entropy
at the conformal points of the boundary. Since  boundary degrees of
freedom localized at the boundary is independent of those in the
bulk in a generic situation,  the boundary renormalization group
flow is independent of bulk renormalization group flow and so the
boundary $g$-function is defined for a non-conformal boundary
condition while the bulk is at the conformal point. In this
subsection we derive holographic g-theorem in NMG from the null
energy condition on matters localized on the bulk boundary $Q$.

Since the bulk in  BCFT   is at conformal point, it is natural to
take $AdS_3$ space as the background geometry for gravity.  One may
take the bulk boundary $Q$ in $AdS_3$ space   as $(r,t , x(r))$ in
the Fefferman-Graham (or $AdS$-kink) coordinates which are related
to Poincare coordinates as $z\equiv e^{-r}$.  In these coordinates
$AdS_3$  space is written as
\[
 ds^2 = L^2\bigg[dr^2 + e^{2A(r)}\Big(-dt^2 + dx^2\Big)\bigg]\,, \qquad A(r) \equiv r\,.
\]
Since one is interested in the renormalization group flow of the
g-function, the bulk boundary $Q$ needs not to be $AdS_2$ space
along the flow except at the conformal points. Therefore, the
profile $x(r)$ of the bulk boundary is taken as a generic function
in this subsection. To address holographic g-theorem in NMG, it is
convenient to use the ADM-decomposed form of $AdS_3$ metric in
Lorentzian signature \footnote{The overall sign of stress tensor
should be changed in this case.}. The above metric of $AdS_3$ space
can be rewritten as the ADM-decomposed form with respect to the bulk
hypersurface $Q$ like the following
\beq \! ds^2=  L^2\bigg[\frac{1}{e^{-2A(r)}+x'(r)^2} d\eta^2  +
e^{2A(r)}\Big\{ - dt^2  +    \Big(e^{-2A(r)}+x'(r)^2\Big)\Big(dr
+\frac{ x'(r)}{e^{-2A(r)}+x'(r)^2}d\eta\Big)^2\Big\}\bigg],
\label{ADM} \eeq
where we have introduced new coordinates $\eta$ as
\[
  \eta \equiv x   -  x(r)\,.
\]
The lapse and shift functions in this  ADM decomposition are read as
\[ N(r) = \frac{L}{\sqrt{e^{-2A(r)}+x'(r)^2}}\,, \qquad N^i  = (N^t,\, N^r) = \Big(0,\,   \frac{x'(r)}{e^{-2A(r)}+x'(r)^2}\Big)\,. \]

Since the above metric form is ADM decomposed one, the indices of
the extrinsic curvature, $K_{ij}$ take only those for surfaces.
Those do not need to take all  bulk indices in our setup.  Note that
the extrinsic curvature of the surface $Q$ in these coordinates is
given by
\bea
 K_{tt} &=& L \frac{e^{2 A(r)} A'(r) x'(r) }{\sqrt{e^{-2
   A(r)}+x'(r)^2  }} \,,  \qquad K_{tr}=0\,,   \nn \\
 K_{rr} &=& -L \frac{ \left(e^{2 A(r)} x'(r)^2+2\right) A'(r) x'(r) +x''(r)}{\sqrt{e^{-2 A(r)}+x'(r)^2 }} \,.
\eea
Null vectors on the surface $Q$ are given by
\beq \xi_{\pm}^{ i} = {e^{-A(r)} \over L}\Big(\pm 1,~ \frac{1}{\sqrt{e^{-2A(r)}+x'(r)^2}}\Big)\,. \label{Nvec}\eeq
Let us recall that  the null energy condition for matters localized on $Q$ with respect to any null vector, $\xi^i_{\pm}$,  is given by
\beq T^{mat}_{ij}~\xi^i_{\pm}\xi^j_{\pm}\ge 0\,.
\eeq
Since $T^{mat}_{ij} = T^{Q}_{ij}$ by boundary condition on $Q$, this
null energy condition can be written as
$T^{Q}_{ij}~\xi^i_{\pm}\xi^j_{\pm}\ge 0$. Now, it is straightforward
to apply this condition  to NMG with boundary terms. Using values of
hatted auxiliary fields $\hat{f}^{i}_{j}$, $\hat{h}^i$, $\hat{s}$
and $\hat{f}$    given in eq.~(\ref{auxvalue}), through a
straightforward calculation, one can see that null energy condition
for null vectors in eq.~(\ref{Nvec}) implies
\beq -\frac{1}{8\pi G L}\bigg[\sigma + \frac{1}{2m^2L^2} \bigg] \frac{e^{-3A(r)} \big(e^{A(r)}x'(r)\big)'}{(e^{-2A(r)}+x'(r)^2)^{3/2}} \ge 0\,. \eeq
Note that this
condition reduces to  the pure Einstein gravity case by taking
$\sigma=1$ and $m^2\rightarrow\infty$.
In this expression one can see that the central charge of dual CFT
appears as an overall coefficient.  

Since we have chosen the Lagrangian
parameter range such that $\sigma + 1/2m^2L^2 >0$ or equivalently
the positive central charge of dual CFT, one obtains the constraint on the profile $x(r)$ as the consequence
of null energy condition
\beq \Big(e^{A(r)}x'(r)\Big)'\le 0 \,. \eeq
This is the essential ingredient for holographic g-theorem and shows
that  the holographic g-theorem still holds in NMG. According to the previous result, it is natural to take the
holographic g-function with respect to the boundary $Q$ specified by
$(r,t, x(r))$  as\footnote{This choice of $g$-function is slightly different from~\cite{Takayanagi:2011zk}.}
\beq \ln g(r) = \frac{c}{6}~ {\rm arcsinh} \Big[ - e^{A(r)}x'(r)\Big]\,, \label{gfunc}
\eeq
where $c$ denotes the central charge of CFT dual to NMG given in
eq.~(\ref{NMGcen}).  This gives us the correct value of boundary
entropy at $x(\infty)=0$ through eq.~(\ref{profx}) and the
coordinate relation. Now, one can see that the constraint on $x(r)$
by null energy condition implies
\beq %
 \frac{\p \ln\, g(r)}{\p r}  \ge 0\,.
\eeq
This is a holographic version of  g-theorem in NMG and reduces to the
ordinary one in the Einstein gravity case.

Though we have considered only the case of $A(r)=r$, which
represents $AdS_3$ space, one may try a more generic function $A(r)$
which represents a $AdS_3$ kink connecting two $AdS_3$ spaces at the
end points. This gravity background corresponds to the simultaneous
renormalization group flow in the bulk and the boundary. Even in
this situation one may ask  the boundary $g$-function can be
introduced and interpreted as the description of boundary degrees of
freedom. One can show that our holographic $g$-function does the job
at least Einstein gravity by elevating the central charge $c$ to the
central charge function $c(r)\sim 1/A'(r)$ in this case.


\section{Boundary entropy and  thermal property}

\subsection{Boundary entropy from   BTZ black holes}
As in the previous section the bulk boundary $Q$ in $AdS_3$ space is
taken as $(\tau, z, x(z))$   in the Poincare  coordinates. The basic
reason for this choice of $Q$ comes from the fact that $Q$ should
preserve $SO(2,1)$ isometry which is manifest in the representation
of $AdS_3$ space by the $AdS_2$ fibration over a line. Since BTZ
black holes are locally isomorphic to $AdS_3$, one may still take
$Q$ in the same way as the case of $AdS_3$ space, which turns out to
be the correct choice. In the following, we consider BTZ black holes
in NMG and obtain boundary entropy from these black holes through
the relation between  thermal partition function and BTZ black
holes. Here, we follow closely the Einstein gravity case.

Let us take the interval of its length $\Delta x$ on the domain  of
the BCFT  ($x <0$)  for BTZ black holes. Using the boundary
conditions on $Q$, it is straightforward to obtain a profile of the
boundary $Q$ in the geometry of BTZ black holes.  Euclideanized BTZ
black holes in Schwarzschild coordinates are represented by
\begin{eqnarray}\label{BTZ}
 ds^2 = \frac{L^2}{z^2} \Big[ f(z) d\tau^2 + \frac{dz^2}{f(z)}+dx^2
 \Big] \,,
\end{eqnarray}
where $f(z) = 1-z^2/z_H^2$ and $z_H$ is a location of the event
horizon. Euclidean time $\tau$ compactified on a circle  satisfies
the periodicity $\tau \sim \tau + 2\pi z_H$ and the temperature of
the dual BCFT is given by $T_{BCFT} = 1/2\pi z_H$.  Through  the
coordinate transformation from $(x,z)$ to $(\eta, z)$ with the
relation $d\eta= dx - x'(z)dz$,  the above  BTZ black hole metric
can  be rewritten in the form of  the ADM-decomposed metric
\begin{eqnarray}\label{met-BTZ}
 ds^2 = \frac{L^2}{z^2} \bigg[ \frac{1}{1+f(z)x'^2(z)}\, d\eta^2+
        f(z)d\tau^2 + \frac{1+f(z)x'^2(z)}{f(z)}
        \Big(dz  + \frac{f(z)x'(z)}{1+f(z)x'^2(z)}d\eta \Big)^2
        \bigg],~~~
\end{eqnarray}
which is more suitable for our purpose.

By introducing an effective (constant) tension on the bulk boundary $Q$ as
\beq
 T_{eff} \equiv \frac{1}{\big( \sigma + \frac{1}{2m^2L^2}  \big)}\, T\,, \eeq
one can see that the boundary condition becomes (see Appendix A.2)
\beq \label{BTZT}
 T_{eff} =  \frac{x'(z)}{L\sqrt{1+f(z) x'(z)^2}}  \,,
\eeq
and $x=x(z)$ satisfies
\begin{eqnarray}\label{profile1}
 \frac{dx}{dz} = \frac{T_{eff} L}{\sqrt{1-T^2_{eff} L^2 f(z)}} \,.
\end{eqnarray}
This result is simply borrowed  from the pure Einstein gravity case
except the replacement of $T$ by $T_{eff}$, since  the metric and
boundary conditions take  the same form with the Einstein case
except the replacement of $T$ by $T_{eff}$. As a result, the profile
of the boundary $Q$ can be described by
\begin{eqnarray}\label{profile2}
 x(z) = z_H {\rm arcsinh} \Bigg( \frac{T_{eff} L }{\sqrt{1-T^2_{eff}L^2}}\,  \frac{z}{z_H}
 \Bigg) \,.
\end{eqnarray}
One can see that this profile inserted in the above ADM-decomposed
metric for BTZ black holes gives us  the representation of those
black holes as $AdS_2$  fibration over a line. One can explicitly
verify that $\eta =const.$  hypersurface is the metric of $AdS_2$,
indeed. Therefore, the bulk boundary $Q$ has $SO(2,1)$ isometry
group as the case of $AdS_3$ space.

On the contrary to the form of the profile $x=x(z)$ which can be
borrowed from Einstein gravity case, the on-shell value of total
action needs to be computed separately in NMG  in order to obtain
HEE of  the interval $A$ with length $\Delta x$ at the asymptotic
boundary. Through a straightforward computation the on-shell value
of the total action is given by (see Appendix A.2)
\begin{eqnarray*}
 S_E &=& S_{bk}+2S_{bd}   =  -\frac{\pi c}{3} \Delta x \cdot T_{BCFT} -\frac{L z_H}{2G}
         \Big( \sigma + \frac{1}{2m^2L^2} \Big)
         \left[ \frac{x(z)}{z^2} \right]^{z_H}_{\epsilon}
         + \frac{c}{6}\frac{z_H \Delta x}{\epsilon^2}  \\
     &=& -\frac{c}{6} \frac{\Delta x}{z_H}
         -\frac{c}{3}\frac{x(z_H)}{z_H}
         + \frac{c}{3}\frac{z_H x(\epsilon)}{\epsilon^2}
         + \frac{c}{6}\frac{z_H \Delta x}{\epsilon^2} \,.
\end{eqnarray*}
To obtain the correct physical result, one needs to be careful by
considering the physical radius at the position $z=\epsilon$. The
asymptotic two dimensional geometry for BCFT at $z=\epsilon$ in
Poincare coordinates is taken by
\beq ds^2_{BCFT} =  f(\epsilon) d\tau^2 + dx^2\,. \eeq
Accordingly, the physical radius of $\tau$ direction on this two
dimensional surface should be taken as
\[ \tilde{z}_H = \sqrt{f(\epsilon)}\, z_H\,. \]
Finally, using $x(z_H)=z_H{\rm arctanh} (T_{eff}\, L)$  and
discarding the divergent parts by appropriate counter terms, one can
see that the action value for the entanglement entropy is given by
\begin{eqnarray}
S_{E}&=& -\frac{\pi c}{6} \Delta x \cdot T_{BCFT}    - \frac{c}{3}
{\rm arctanh}  (T_{eff}\, L) \,,
\end{eqnarray}
where $T_{BCFT}=1/2\pi z_H$.
Note that the final form of total Euclidean action value in this
case is identical with the one in Einstein gravity except the
adaptation of central charge relevant to NMG.

In  the holographic construction standard thermal partition function
is given at the leading order by  the on-shell value of Euclidean
gravity action with temporal length $\beta$
%
%
%
\beq Z^{BCFT}_{th}(\beta) = e^{-S_{E}(\beta)}\,. \eeq
This thermal partition function also takes the same form with the
Einstein gravity case though the central charge is different.
Therefore, one can see that the various properties inherited from
$S_{E}$  are same with the Einstein gravity case.   At a high
temperature the BCFT partition function, which is  defined on
cylinder of the length $\Delta x$ with two boundaries $\alpha,
\beta$, is given by $Z_{\alpha \beta} \simeq g_{\alpha}g_{\beta}
e^{-E_0\Delta x}$   and so the boundary entropy can be also read in
this approach.  As far as BTZ black holes are dominant at high
temperatures as in Einstein gravity, the same result about boundary
entropy with eq.~(\ref{bdent}) is obtained in this approach, too. To
show the thermal phase in NMG with the bulk boundary $Q$ or the BTZ
black hole dominance at high temperatures, thermal $AdS_3$ are
studied in the next section.

The  holographic thermal entropy from BTZ black holes can be read from
\beq {\cal S}_{th} = - S_{E} + \beta \frac{\p}{\p \beta} S_{E}\,.
\eeq
In our case $S_{E}$ is given by $S_{E} = S_{bk} +2 S_{bd}$.  Since
$S_{bk} \sim 1/\beta$ and $S_{bd}$ is independent of $\beta$,  one
can see that  ${\cal S}_{th} = -2S_{bk}-2S_{bd}$.  As a result, thermal entropy
for BTZ black holes for the system of length $\Delta x$ is given by
\beq
 {\cal S}_{th} = \frac{\pi c}{3} \Delta x \cdot T_{BCFT}
      + \frac{c}{3} {\rm arctanh}  (T_{eff}\, L) \,.
\eeq
%

\subsection{Holographic thermal property of $AdS$ black holes}
The metrics  of thermal solitons and non-rotating new type black holes \cite{Oliva:2009ip} are given by
\beq
 ds^2 = \frac{L^2}{z^2}\bigg[d\tau^2 +  \frac{1}{h(z)} dz^2 + h(z)  dx^2\bigg]\,.  \qquad h(z) = 1 + B z + C z^2\,,
\eeq
\beq
ds^2 = \frac{L^2}{z^2}\bigg[ f(z)d\tau^2 +  \frac{1}{f(z)}dz^2  + dx^2\bigg]\,, \qquad f(z) = 1 +b z + cz^2\,,
\eeq
Let us focus on $B=b=0$ case which corresponds to the case of non-rotating BTZ black holes and the corresponding thermal solitons.
%
Through  the coordinate transformation from $(x,z)$ to $(\eta, z)$
with the relation $d\eta= dx - x'(z)dz$,  the thermal AdS  metric
can  be rewritten in the form of  the ADM-decomposed metric
\begin{eqnarray}\label{met-sol}
 ds^2 = \frac{L^2}{z^2} \bigg[ \frac{h(z)}{1+h(z)^2 x'(z)^2}\, d\eta^2+
        d\tau^2 + \frac{1+h(z)^2 x'(z)^2 }{h(z)}
        \Big(dz  + \frac{h(z)^2 x'(z)}{1+h(z)^2 x'(z)^2}d\eta \Big)^2
        \bigg].~~~~~
\end{eqnarray}
From the boundary condition, the profile of $Q$ can be found (see
Appendix.A.3) . By introducing an effective (constant) tension on
the bulk boundary $Q$ as
\beq
 T_{eff} \equiv \frac{1}{\big( \sigma + \frac{1}{2m^2L^2}  \big)}\, T\,, \eeq
one can see that the boundary condition becomes
\[
 T_{eff} = \frac{h(z)^2 x'(z)}{L \sqrt{h(z)(1+h(z)^2 \,x'(z)^2)}}  \,,
\]
and $x=x(z)$ satisfies
\begin{eqnarray}\label{profile1}
 \frac{dx(z)}{dz} =\frac{L \,T_{eff}}{h(z) \sqrt{h(z)-L^2 T_{eff}^2}} \,.
\end{eqnarray}
This result is the same as the pure Einstein gravity case except the replacement of $T$ by $T_{eff}$. 
Therefore, the profile of the boundary $Q$ can be described by
\begin{eqnarray}\label{profile2}
 x(z) = z_0 \arctan \Bigg( {  {L\,T_{eff} \, z}  \over {z_0 \sqrt{h(z)-L^2 T_{eff}^2}}  }  \Bigg) \,.
\end{eqnarray}
%
The on-shell value of the total action is given by (see
Appendix.A.3)
\begin{eqnarray}
 S_E  =   {{L z_H} \over {2G}}\Big( \sigma +{1 \over {2 m^2 L^2}} \Big)
         \left[ -{ \pi \over {2 z_0}}+{1 \over \epsilon^2}
\left(x(\epsilon)+{{\pi z_0} \over 2}  \right) -{{x(\epsilon)} \over
z_0^2} \right] \,.
\end{eqnarray}
%

The smoothness of the geometry near $z=z_0$  gives us  the
periodicity of $x$ as $x\sim x + 2\pi z_0$, which means that the
radius of $x$ direction is $z_0$. As in the case of BTZ black holes,
the asymptotic two dimensional geometry for BCFT at $z=\epsilon$ is
taken as \beq ds^2 =   d\tau^2 + h(\epsilon) dx^2\,. \eeq
By matching  the physical radius of $x$ as
\[ \tilde{z}_0 = \sqrt{h(\epsilon)}\, z_0\,, \]
and discarding the divergent part by appropriate
counter terms, one can see that the quantity relevant to the
entanglement entropy is given by
\beq S_E = -   \Big( \sigma +{1 \over {2 m^2 L^2}} \Big) {{ \pi L
z_H} \over {8 G z_0} } = - {\pi \over 24} {c \over {\Delta x \,\,
T_{BCFT} }} \,, \eeq where $T_{BCFT} = 1/(2 \pi z_H) $.
This also takes the same form as the Einstein case  and it leads to
the same boundary entropy in NMG as the one in Einstein gravity
except scaling of Newton's constant and cosmological constant
through central charge. Since all the expression is identical with
the Einstein gravity case, thermal properties are same with the
Einstein gravity case. For instance,  thermal solitons are preferred
at low temperatures and BTZ  black holes at higher temperatures, of
which phase transition point is given by $z_0=z_H$.


Next, let us consider new type black hole
case~\cite{Bergshoeff:2009aq} in NMG, which exists only at the
special parameters given by $\sigma=1$, and $2 m^2 L^2=1$.
%
%
Euclideanized new type black holes in Schwarzschild coordinates can
be represented by (\ref{BTZ}) with $f(z) =(1-z/z_H)(1-z/z_{in})$ and
$z_H$ is a location of the event horizon and $z_{in}$ is a certain
parameter in new type black holes~\cite{Oliva:2009ip}. Euclidean
time $\tau$ compactified on a circle satisfies  the periodicity
$\tau \sim \tau + 4\pi ({1 \over z_H}-{1 \over z_{in}})^{-1} $ and
the temperature of the dual BCFT is given by $T_{BCFT} = {1 \over {4\pi}} ({1 \over z_H}-{1 \over z_{in}})$.
 Through  the coordinate transformation from $(x,z)$ to $(\eta, z)$ with the relation $d\eta=
dx - x'(z)dz$, the  metric  (\ref{BTZ})  can be rewritten in the form of the
ADM-decomposed metric (\ref{met-BTZ}).

Even though the metric takes the same form with the BTZ black hole
case, there are some definite differences between them, taking  into account
boundary entropy. In fact, some issues described in the following
should be resolved in new type
black holes. 
%
Specifically, as mentioned before, $R^{\eta}_{i}$  vanishes   for
BTZ black holes and the charge conservation of Brown-York tensor holds from
the second Gauss-Codacci equation. However, for new type black holes
one can see that $R^{\eta}_{i}$ does not vanish as follows:
\begin{equation}
R^{\eta}_z = \frac{\left(2-2 f(z)+z f'(z)\right)}{2 L^2}   x'(z) \,.
\end{equation}
%
Note that BTZ black holes satisfy the relation $ 2-2 f(z)+z
f'(z)=0$. Moreover,  charges of Brown-York tensor for new type black
holes are not conserved in general (see Appendix.B).  This
phenomenon is attributed to the existence of a new parameter in the
metric of (static) new type black holes  which makes their
geometrical and physical properties different from those of the BTZ
black holes. The charge conservation of Brown-York tensor is a usual
assumption in $AdS/CFT$ correspondence, which is related to the
conformal invariance in the CFT side.   However, it is not so clear
in NMG that the conservation of Brown-York tensors is requirement or
relevant assumption.

For new type black holes there is no profile of $Q$ to satisfy
the boundary condition (\ref{boundary}) with a constant tension $T$.
One may consider a more generic matter Lagrangian to find the profile 
of $Q$ from the boundary condition. But in these cases the situation 
would be different from the simplest case we considered in this paper. It 
is expected that the charge conservation for Brown-York tensor 
would not be satisfied and the surface $Q$ would not be $AdS_2$, in 
contrast to BTZ black hole case. Therefore $AdS/BCFT$ for new type black 
holes should be separately investigated further.
%
%
%
This is because of the property of new
type black holes not locally isomorphic to $AdS_3$, contrary to
BTZ black hole case. So, for new type black holes,
one might need to take a different ansatz for $Q$ to satisfy
charge conservation and boundary condition.


%
%


Even with the  problem of specifying the surface $Q$, however, it is expected
that the new type black holes with bulk boundary have a similar
thermal property to BTZ black holes with bulk boundary.



%
%


\section{A simple derivation}
In this section we rederive our results presented in the previous
sections through a few simple steps. First, let us recall that the
radius $L$ of  AdS space in NMG is related to the parameters
$\sigma, m$ and $ l$ in the Lagrangian~(\ref{Action}) by
\beq
    \frac{1}{\ell^2} = \frac{1}{L^2}\Big(\sigma - \frac{1}{4m^2L^2}\Big)\,.
\eeq
Second, one may notice that the auxiliary field formalism in some
kinds of higher derivative theories is particularly appropriate  for
$AdS/CFT$ correspondence. After the construction of
three-dimensional NMG action, there are some studies on critical and
non-critical gravities with higher derivatives in various
dimensions.  One of the interesting observations in these theories
on AdS space is the possibility of the consistent truncation for
unwanted ghost-like modes by appropriate boundary conditions. It was
argued that this consistent truncation in four-dimensional case may
be elevated  to the full non-linear level~\cite{Maldacena:2011mk}
and indeed shown that is the case  through symmetric auxiliary
fields~\cite{Hyun:2011ej}  (see also~\cite{Metsaev}). In the following we propose how to extend this effective description of specific higher curvature gravity to NMG even with a bulk boundary.

By taking the same procedure in the case of  four dimensional
non-critical Einstein-Weyl gravity~\cite{Hyun:2011ej}, one may lift
the value in AdS space  to the full metric field  for the auxiliary
tensor field $f_{\mu\nu}$. Explicitly, this means that one can take
the auxiliary field $f_{\mu\nu}$ as
\beq f_{\mu\nu} = -\frac{1}{m^2L^2}g_{\mu\nu}\,,  \qquad  f = -\frac{3}{m^2L^2}\,, \eeq
where $g_{\mu\nu}$ denotes  the fluctuating metric field not the
metric value in AdS space. By inserting this ansatz to two tensor
form of NMG bulk action given in~(\ref{Action}), one obtains
\beq \CL_{NMG} = \frac{1}{16\pi G}\bigg[\sigma R + \frac{2}{\ell^2}
+ \frac{1}{2m^2L^2} R  +\frac{3}{2m^2L^2}\bigg] = \frac{1}{16\pi G}
\Big(\sigma + \frac{1}{2m^2L^2}\Big)\bigg[R + \frac{2}{L^2}\bigg]\,.
\eeq
Note that this is the  just action for the pure  Einstein gravity
with rescaled Newton's constant $G_{eff}\equiv G/(\sigma +
1/2m^2L^2)$ and with the substituted  cosmological constant, $L$.

This simple computation explains  various results about NMG in the
setup of $AdS/CFT$ correspondence. For example, the central charge
of dual CFT for NMG is given simply by the rescaling of Newton's
constant  and the substitution of cosmological constant as
\[ c = \frac{3L}{2G} \Big(\sigma + \frac{1}{2m^2L^2}\Big)\,. \]

Moreover, one can see that the GH term in NMG is also reduced to the
one in the pure Einstein gravity by the same lift of auxiliary field
in the boundary. To see this, note that the relevant auxiliary
fields in NMG GH term are taken by
\beq \hat{f}_{ij} = -\frac{1}{m^2L^2}\gamma_{ij}\,, \qquad \hat{f} = -\frac{2}{m^2L^2}\,. \eeq
Then, the GH term~(\ref{GH}) becomes
\beq S^{NMG}_{GH} = \frac{1}{16\pi G} \int d^2x\sqrt{-\gamma}
\Big[2\sigma K - \frac{1}{m^2L^2}K + \frac{2}{m^2L^2}K \Big] =
\frac{1}{8\pi G_{eff}} \int d^2x\sqrt{-\gamma}~   K\,,\eeq
which shows us that NMG GH term also reduced to the one in the pure
Einstein gravity with the same rescaling of Newton's constant.
Another point to address in the boundary entropy is the existence of
matter sector on the boundary $Q$, which is described, in the
simplest case,  by
\[
 S^{mat}_{Q} =   -  \frac{1}{8\pi G} \int_Q \sqrt{\gamma}~  T\,.
\]
Since the GH term is reduced to the pure Einstein case with
effective Newton's constant $G_{eff}$, it is natural to rewrite the
matter part in terms of $G'$ as
\beq S^{mat}_{Q} =   -  \frac{1}{8\pi G_{eff}} \int_Q
\sqrt{\gamma}\, T_{eff}\,, \qquad T_{eff} \equiv
\frac{1}{\big(\sigma + \frac{1}{2m^2L^2}\big)}\, T\,. \eeq
Hence, the boundary action on $Q$ becomes
\beq S_{Q} +S^{mat}_{Q}=  \frac{1}{8\pi G_{eff}}\int_Q d^2x
\sqrt{\gamma} \Big[ K - T_{eff}\Big]\,. \eeq
This is  our  result for the effective description of NMG with a bulk boundary in the context of the AdS/BCFT correspondence.

The above short computation reveals that the effective description
of NMG in the setup of $AdS/CFT$ correspondence is given simply by
the pure Einstein gravity with some rescaling and substitution. Even
with boundary, the above effective description becomes unchanged. As
a result, most of  our computation in previous sections are
understood as simple rescaling of Newton's constant and the
substitution of the cosmological constant.  This simple derivation
holds for BTZ black holes and other cases except new type  black
holes.

One can consider this simple derivation as another check of our
results given in the previous sections.  Some comments may clarify
the meaning of this derivation  since the role of higher derivatives
may be somewhat  different according to the spacetime dimensions.
Note that the above derivation  is based on the possibility of
consistent truncation. In the case of four-dimensional  conformal or
noncritical Einstein-Weyl gravity, this truncation is achieved by
appropriate boundary conditions at the asymptotic infinity since the
unwanted ghost-like massive modes fall off more slowly than $\log$
or massless  modes.  On the contrary, three-dimensional NMG is
regarded as the covariant completion of Pauli-Fierz massive graviton
theory, and  so  propagating physical massive graviton modes are
allowed with an appropriate choice of Lagrangian parameters.
Accordingly, one may worry about the inappropriate truncation of
physical modes  in the NMG case. However, one may recall that the
unitarity of graviton modes and dual CFT is incompatible in the case
of the three-dimensional Einstein gravity and even in the NMG
case~\cite{Bergshoeff:2009aq}. All our results in the previous
sections belong to the parameter regions $c>0$ where dual CFT is
unitary or  BTZ black holes have positive mass while the gravitons
are ghost-like modes and can be truncated for $m^2 > 0$.  Therefore,
gravitons may be truncated consistently in our setup of $AdS/CFT$
correspondence which requires dual CFT be unitary. More explicitly,
our boundary conditions taken at the asymptotic infinity are those
of Brown-Henneaux~\cite{Brown:1986nw} which excludes new type black
holes. This explains the above derivation does not apply to those
black holes. See Appendix B for some attempts of $AdS/BCFT$
correspondence  in new type black holes.


\section{Conclusion}

It is natural to ask higher curvature effects on holographic
description of $BCFT$ after its construction for Einstein gravity.
As in HEE, $AdS/BCFT$ correspondence has no generic prescription for
higher curvature effects, and so we have explored one  simple
extension of the original proposal of $AdS/BCFT$ correspondence on
Einstein gravity by considering the correspondence for the so-called
NMG on three dimensions. NMG is most natural setting to see higher
curvature effects in the sense that it is unique curvature square
gravity consistent with holographic c-theorem.

In this paper we have followed closely the construction in the
Einstein gravity case and  computed  in three different ways  the
boundary entropy for BCFT defined on a half space or its
conformally-related disk. Concretely, boundary entropy is obtained
through its relation with disk amplitude, HEE and black holes. We
showed that all approaches give us the same result for boundary
entropy and it is simply given by the same form with the Einstein
gravity case except the fact that the relevant central charge is
taken for NMG case not for the pure Einstein gravity one.  By using
auxiliary field formulation of NMG, we have also showed that the
holographic g-theorem still holds in NMG.   Thermal properties of
$BCFT$ are also studied holographically in NMG.  Furthermore,
we have showed that these results can be understood by the effective
description of NMG by the auxiliary field which is recently analyzed
and advocated  in the case of non-critical Einstein gravity in four
dimensions.

Our study is just  a small step toward understanding of  higher
curvature effects on $AdS/BCFT$ correspondence. One immediate
question one may ask is about effects by even higher curvature
terms.  There are several extensions of NMG consistent with
holographic c-theorem with even more higher curvature terms and it
is interesting to see their effects on the correspondence. Most
natural expectation is that only the central charge will be modified
and all the results will be same with Einstein gravity except the
substitution in various expression by the modified central charge.
The counter terms for the divergence cancelation should also be
addressed properly in the case of BCFT since the counter terms are
more subtle with higher curvature terms as shown
in~\cite{Kwon:2011jz}. Another interesting question is the higher
curvature effects on  higher dimensional spacetime. In this case
Lovelock gravity may be more adequate for the analysis as in HEE
studies.

It needs to be mentioned that there may be $\log$-modes in higher
derivative theories at a specific value of higher derivative
couplings, which is beyond the scope of this paper. As can be seen
from the fact that the simple derivation through auxiliary field
formalism depends on the truncation   of slow fall-off modes
including $\log$ ones, the original proposal for $AdS/BCFT$
correspondence in Einstein gravity should be modified when these
slow fall-off modes in higher derivative gravity are included.
Specifically in the case of $AdS_3/BCFT_2$, this means that modified
construction needs to be considered for the more general fall-off
boundary conditions than the Brown-Henneaux ones.  At the specific
coupling in NMG required for the existence of $\log$-modes, the
central charge vanishes and the theory is argued to be null under
the strict Brown-Henneaux boundary conditions, which is consistent
with our results about the zero  g-function value at that coupling.
However,  the boundary entropy of dual CFT may be non zero with
$\log$-modes,  since $\log$ modes are related to non-unitary
$\log$-CFT according to $\log$-gravity conjecture and it is
plausible  to envisage some non-vanishing boundary entropy in this
case.

\section*{Acknowledgements}


S.N and S.H.Y were supported by the National Research Foundation of
Korea(NRF) grant funded by the Korea government(MEST) through the
Center for Quantum Spacetime(CQUeST) of Sogang University with grant
number 2005-0049409. S.H.Y would like to thank  Seungjoon Hyun,
Jaehoon Jeong and Wooje Jang at Yonsei University for some
discussion. S.N and J.D.P were supported by a grant from the Kyung
Hee University in 2009(KHU-20110060). S.N was supported by Basic
Science Research Program through the National Research Foundation of
Korea(NRF) funded by the Ministry of Education, Science and
Technology(No.2011-0004328).  Y.K   was supported by the National
Research Foundation of Korea Grant funded by the Korean
Government(Ministry of Education, Science and Technology)
(NRF-2011-355-C00027).

\newpage
\appendix
 \renewcommand{\theequation}{A.\arabic{equation}}
  \setcounter{equation}{0}
\section{Some formulae for on-shell action value of NMG}
In this appendix, we present  some computational details    to
obtain boundary entropy, {\it i.e.} $\ln g$, associated with a
boundary $\p M$. In our convention, Euclidean NMG actions with
boundary terms on $Q$, while omitting asymptotic boundary term on
$M$, are  taken as Eq.~(\ref{Eaction}) which can be written as
\beq
 S_{bk} \equiv S =  -\frac{1}{16\pi G} \int_N \sqrt{g}~ \CL_{bk}\,, \qquad
S_{bd} \equiv S_Q  +
      S^{mat}_{Q} =  -\frac{1}{8\pi G} \int_Q \sqrt{\gamma}~ \CL_{bd}  \label{Action-s}\,.
\eeq

\subsection{Disk amplitude}
The following are some steps to obtain the on-shell value of
Euclidean action given in eq.~(\ref{diskamp}) in order for the
boundary entropy as the disk amplitude:
   Let us first consider the
coordinates transformation from $(\tau, x, z)$ to $(r, \theta,
\phi)$ in accordance with
\bea \tau = r\sin\theta \cos \phi\,, ~~
          x = r\sin\theta \sin \phi\,, ~~
         z =  r_D\sinh\frac{\rho_*}{L}  + r\cos \theta\,. \nn
\eea
Then, the metric (\ref{metric}) becomes
\bea\label{metric-disk}
 ds^2 = L^2 \frac{dr^2 + r^2 d\theta^2 + r^2 \sin^2\theta d\phi^2}{(r\cos\theta+r_D \sinh\frac{\rho_*}{L})^2}
\eea
and surface integration does
\[\left.
  \int_Q  \sqrt{\gamma} = \int d\theta d\phi~
  \frac{L^2 r^2 \sin\theta}{(r\cos\theta + r_D \sinh\frac{\rho_*}{L})^2}
  \right|_{r=r_D \cosh\frac{\rho_*}{L}}\,.
\]
\\
Using the above metric (\ref{metric-disk}) and applying the boundary
condition~$(\ref{boundary})$, one can see that integrands for the
bulk and boundary parts become
\[
   {\cal L}_{bk}
   = -\frac{4}{L^2} \Big( \sigma + \frac{1}{2m^2L^2} \Big) \,,   \qquad
   {\cal L}_{bd}
   = \frac{r_D \sinh\frac{\rho_*}{L}}{L}
   \Big( \sigma + \frac{1}{2m^2L^2} \Big) \frac{1}{r} \,,
\]
where we used the tension given by
\begin{eqnarray}\label{tension}
 T = \frac{r_D\sinh\frac{\rho_*}{L}}{L}
    \Big( \sigma + \frac{1}{2m^2L^2} \Big) \frac{1}{r} \,.
\end{eqnarray}
Some values related to auxiliary fields $f^{\mu\nu}$ are as follows;
\begin{eqnarray*}
 && \hat{s} = -\frac{1}{m^2 L^2}\,, ~~~~~
 \hat{f} = -\frac{2}{m^2 L^2}\,, ~~~~~
 D_r \hat{f} = 0\,, \\
 && \frac{1}{2} \gamma_{ij}D_r\hat{f}^{ij}
 = \frac{r_D \sinh\frac{\rho_*}{L}}{m^2L^3}
 \cdot \frac{2}{r}  \,, ~~~~~
 K_{ij}\hat{f}^{ij} = -\frac{r_D \sinh\frac{\rho_*}{L}}{m^2L^3}
 \cdot \frac{2}{r}\,.
\end{eqnarray*}
As a result, the on-shell value of total Euclidean action for the disk are given by
\begin{eqnarray*}
 S_{E} &=& S_{bk} + S_{bd}  \\
    &=& \frac{1}{4\pi G L^2} \Big( \sigma
    + \frac{1}{2m^2L^2} \Big) \int^{r_0}_{\epsilon} \frac{L^3}{z^3}dz
    \int^{r_*}_{0} 2\pi r dr    \\
    && \left. -\frac{1}{8\pi G} \int^{\theta_0}_0 d\theta \int^{2\pi}_{0} d\phi
    \frac{L^2 r^2 \sin\theta}{(r\cos\theta + r_D \sinh\frac{\rho_*}{L})^2}
    \cdot \frac{r_D \sinh\frac{\rho_*}{L}}{L}
    \Big( \sigma + \frac{1}{2m^2L^2} \Big) \frac{1}{r} \right|_{r=r_D
    \cosh\frac{\rho_*}{L}} \\
   &=& \frac{L}{4G} \Big( \sigma + \frac{1}{2m^2L^2} \Big)
    \bigg[ \frac{r_D^2}{2\epsilon^2} + \frac{r_D \sinh\frac{\rho_*}{L}}{\epsilon}
    + \log \Big( \frac{\epsilon}{r_D} \Big) - \frac{1}{2} - \frac{\rho_*}{L}
    \bigg] \,,
\end{eqnarray*}
where $\epsilon$ is introduced for the UV cutoff. Parameters $r_0$,
$r_*$ and $\theta_0$ for the integration ranges are given by
\begin{eqnarray*}
 r_0 = r_D e^{\frac{\rho_*}{L}} \,,~~~
  r_*^2 = r_D^2 \cosh^2 \frac{\rho_*}{L} - (z-r_D\sinh
  \frac{\rho_*}{L})^2  \,, ~~~
  \theta_0 = -{\rm arccos}\Big(\tanh\frac{\rho_*}{L}\Big)\,.
\end{eqnarray*}
%

\subsection{BTZ black hole}

%
Using the ADM decomposition, the BTZ black hole metric can be
represented by eq.(\ref{met-BTZ}) with $f(z)=1-z^2/z_H^2$. The lapse
function $N(z)$ and shift vector $N^i$ are given by
\begin{eqnarray*}
 N(z) = \frac{L}{z\sqrt{1+f(z)x'(z)^2}} \,, ~~~~~
   N^i = (N^{\tau}, N^z) = \Big( 0,\frac{f(z)x'(z)}{1+f(z)x'(z)^2}
   \Big) \,.
\end{eqnarray*}
For a space-like unit vector $n^{\mu}$ normal to the hypersurface
$Q$,  the induced metric $\gamma_{\mu\nu}$ are introduced as
$\gamma_{\mu\nu} = g_{\mu\nu}-n_{\mu}n_{\nu}$. In our case
$n^{\mu}$ is given by
\begin{eqnarray}
 n^{\mu} = \frac{z}{L\sqrt{1+f(z)x'(z)^2}}(0, -x'(z)f(z),1) \,.
\end{eqnarray}
The induced metric on the surface $Q$ in a matrix form is given by
\begin{eqnarray}
 \gamma_{\tau\tau} = \frac{L^2 f(z)}{z^2}\,, ~~~
   \gamma_{zz} = \frac{L^2(1+f(z) x'(z)^2)}{z^2f(z)} \,, ~~~
     \gamma_{\tau z} =0 \,.
\end{eqnarray}
Then, the extrinsic curvature defined by (\ref{ext-def}) is given by
\begin{eqnarray}
\hspace{-0.5cm}
  K_{\tau\tau} =  \frac{L f(z) x'(z)}{z^2\sqrt{1+f(z) x'(z)^2}} \,,
  ~
  K_{zz} = \frac{L[x'(z)+f(z)(f(z) x'(z)^3-z
  x''(z))]}{z^2f(z)\sqrt{1+f(z)x'(z)^2}} \,, ~
  K_{\tau z} =0 \,.
\end{eqnarray}
After all, the boundary condition gives us  the relation for
the profile of the surface $Q$ as
\begin{eqnarray} \label{BTZTT}
 T = \Big( \sigma + \frac{1}{2m^2L^2}  \Big)
 \frac{x'(z)}{L\sqrt{1+f(z)x'(z)^2}}  \,.
\end{eqnarray}
Some values of hatted quantities  relevant  to the computation  are
\begin{eqnarray*}
 && \hat{s} = -\frac{1}{m^2L^2} \,, \qquad \hat{h}^i = 0 \,,  \qquad \hat{f} =
 -\frac{2}{m^2L^2}  \,,  \\
 && \hat{f}^{\tau\tau} = -\frac{z^2}{m^2L^4}\frac{1}{f(z)}\,, \qquad
 \hat{f}^{zz} = -\frac{z^2}{m^2L^4}\frac{f(z)}{(1+f(z)x'(z)^2)}   \,.
\end{eqnarray*}
The integrands for the bulk action, the boundary one and the
extrinsic curvature scalar in the BTZ black hole case are given by
\begin{eqnarray}
  {\cal L}_{bk} =
       -\frac{4}{L^2} \Big( \sigma + \frac{1}{2m^2L^2} \Big) \,, ~~
  {\cal L}_{bd}
       = \Big( \sigma + \frac{1}{2m^2L^2} \Big)K - T   \,, ~~
   K = \frac{2x'(z)}{L\sqrt{1+f(z)x'(z)^2}} \,. \label{BTZav}
\end{eqnarray}
Using above quantities,integrands and extrinsic curvature scalar,
the on-shell values of the action can be obtained. Firstly, the
value of bulk action is given by
\begin{eqnarray}
 S_{bk}  &=& -\frac{1}{16\pi G} \int d\tau dz dx \frac{L^3}{z^3}
          \bigg[-\frac{4}{L^2}\Big( \sigma + \frac{1}{2m^2L^2} \Big)
          \bigg]
          \nn   \\
       &=& -\frac{c}{6}\frac{\Delta x}{z_H}
          + \frac{L z_H}{G}\Big( \sigma + \frac{1}{2m^2L^2} \Big)
          \int^{z_H}_{\epsilon} \frac{dz}{z^3} \cdot x(z)
          + \frac{c}{6}\frac{z_H \Delta x}{\epsilon^2} \,.
\end{eqnarray}
And the action value for the boundary part becomes
\begin{eqnarray}
 S_{bd}
     &=& -\frac{L z_H}{4G} \Big( \sigma + \frac{1}{2m^2L^2} \Big)
         \int^{z_H}_{\epsilon} \frac{dz}{z^2} \cdot x'(z)  \nn \\
     &=& -\frac{L z_H}{4G} \Big( \sigma + \frac{1}{2m^2L^2} \Big)
         \left[ \frac{x(z)}{z^2} \right]^{z_H}_{\epsilon}
         -\frac{L z_H}{2G} \Big( \sigma + \frac{1}{2m^2L^2} \Big)
         \int^{z_H}_{\epsilon} \frac{dz}{z^3} \cdot x(z) \,.
\end{eqnarray}
Finally, the on-shell value of total Euclidean action for the BTZ
black hole case is given by
\begin{eqnarray}
 S_E = S_{bk}+2S_{bd}   =
     -\frac{c}{6} \frac{\Delta x}{z_H}
         -\frac{c}{3}\frac{x(z_H)}{z_H}
         + \frac{c}{3}\frac{z_H x(\epsilon)}{\epsilon^2}
         + \frac{c}{6}\frac{z_H \Delta x}{\epsilon^2} \,.
\end{eqnarray}

\subsection{Thermal $AdS_3$ or thermal solitons}
The metric of thermal  solitons  can  be written in the form of  the
ADM-decomposed one for the bulk boundary $Q$ as eq.(\ref{met-sol})
with $h(z)=1-z^2/z_0^2$. Lapse function and shift vector can be seen
as
\begin{eqnarray}
 N=\frac{L h(z)}{z\sqrt{h(z)(1+h(z)^2x'(z)^2)}}\,,~~~~
   N^i = (N^{\tau}\,, N^z) = \Big(0\,, \frac{h(z)^2x'(z)}{1+h(z)^2x'(z)^2}\Big)
   \,.
\end{eqnarray}
According to the same procedure for the BTZ black hole case, we can
obtain the induced metric for the thermal solitons. Choosing a
space-like unit vector normal to $Q$
\begin{eqnarray}
 n^{\mu} = \frac{z}{L\sqrt{h(z)(1+h(z)^2x'(z)^2)}}(0\,, -h(z)^2x'(z)\,, 1) \,.
\end{eqnarray}
Then, induced metric $\gamma_{ij}$ on $Q$ can be given in the form
of
\begin{eqnarray}
 \gamma_{\tau\tau} = \frac{L^2}{z^2} \,, ~~~
   \gamma_{zz} =  \frac{L^2(1+h(z)^2x'(z)^2)}{z^2 h(z)} \,, ~~~
     \gamma_{\tau z} = 0 \,.
\end{eqnarray}
Now, using eq.(\ref{ext-def}) with the above induced metric, the
extrinsic curvature can be read as
\begin{eqnarray}
 K_{\tau\tau }
    &=& \frac{L h(z)^2 x'(z)}{z^2\sqrt{h(z)(1+h(z)^2x'(z)^2)}} \,,
     ~~~ K_{\tau z} = 0\\
 K_{zz} &=& \frac{L[x'(z)(1+h(z)^2x'(z)^2)
         -z(h(z) x''(z)+h'(z)
         x'(z))]}{z^2\sqrt{h(z)(1+h(z)^2x'(z)^2)}} \,.
\end{eqnarray}
Adopting the boundary condition on the bulk boundary  $Q$,  one
finally obtains
\begin{eqnarray}
  T=\Big(\sigma+\frac{1}{2m^2L^2}\Big)\frac{h(z)^2x'(z)}{L\sqrt{h(z)(1+h(z)^2x'(z)^2)}}
  \,.
\end{eqnarray}
Some values of hatted quantities in this case are given by
\begin{eqnarray*}
 && \hat{s} = -\frac{1}{m^2L^2} \,, \qquad  \hat{h}^i = 0 \,,  \qquad \hat{f} =
 -\frac{2}{m^2L^2}  \,,  \\
 && \hat{f}^{\tau\tau} = -\frac{z^2}{m^2L^4}\,, \qquad
 \hat{f}^{zz} = -\frac{z^2h(z)}{m^2L^4(1+h(z)^2x'(z)^2)}
 \,,
\end{eqnarray*}
and extrinsic curvature scalar is
\begin{eqnarray}
 K = \frac{2h(z)^2x'(z)}{L\sqrt{h(z)(1+h(z)^2x'(z)^2)}} \,.
\end{eqnarray}
The Euclidean action is given by
\begin{eqnarray}
 S_{bk} &=&  { {L} \over {4\pi G}} \Big( \sigma +{1 \over {2 m^2 L^2}} \Big)
         \int d\tau dz dx {1 \over z^3}   \nn \\
    &=& {{L z_H} \over  G} \Big( \sigma +{1 \over {2 m^2 L^2}} \Big)
         \left[ \int_{\epsilon}^{z_{\ast}} dz {1 \over z^3}
         \left(x(z)+{{ \pi z_0} \over 2} \right)
         +\int_{z_\ast} ^{z_0} dz {{\pi z_0} \over z^3} \right] \\
S_{bd}
     &=&  - {{L z_H} \over {2 G}}\Big( \sigma +{1 \over {2m^2 L^2}} \Big)
         \int_{\epsilon} ^{z_\ast} dz {1 \over z^2} h(z) x'(z)  \,.
\end{eqnarray}
The on-shell value of the total action
is given by
\begin{eqnarray}
 S_E = S_{bk}+ S_{bd} =
         {{L z_H} \over {2G}}\Big( \sigma +{1 \over {2 m^2 L^2}} \Big)
           \left[ -{ \pi \over {2 z_0}}+{1 \over \epsilon^2}
           \left(x(\epsilon)+{{\pi z_0} \over 2}  \right)
           -{{x(\epsilon)} \over z_0^2} \right] \,.
\end{eqnarray}
%


 \renewcommand{\theequation}{B.\arabic{equation}}
  \setcounter{equation}{0}
\section{New type black holes and HEE}
%
%


In this section, let us consider new type black holes in NMG.
Euclideanized new type black holes can be recast in the form of the
ADM-decomposed metric (\ref{met-BTZ}) with $f(z)
=(1-z/z_H)(1-z/z_{in})$.
%
%
First, consider action values (\ref{Action-s}) for new type black
holes. Note that both BTZ black hole and new type black hole have
the following relation,
\begin{eqnarray} \label{rel}
 f^{n}(z)&=&0  \qquad \qquad (n \ge 3) \,,\nonumber  \\
 z^2 f''(z) +{2(f(z)-z f'(z)-1)} &=& 0 \,.
\end{eqnarray}
Using these relation, the integrands of bulk action and boundary
action become
\begin{eqnarray}
{\cal L}_{bk} &=& \Big(\sigma +\frac{1}{2 m^2  L^2}\Big) \frac{2 z f'(z)-4 f(z)}{L^2}  \,, \\
{\cal L}_{bd} &=&
   \Big(\sigma +\frac{1}{2 m^2 L^2 }\Big) K
    +  \frac{(2 f(z)-z f'(z)-2) (2 f(z)-z f'(z)) f(z) x'(z)^3}{4 m^2 L^3 (1+f(z)\, x'(z)^2)^{3/2}}
    - T \,,
\end{eqnarray}
where
\begin{equation*}
K= \frac{[ f(z) (4 f(z)-z f'(z)) x'(z)^3+(4 f(z)-2 z f'(z)) x'(z)-2
z f(z) x''(z) ]}{2 L(1+f(z) \,x'(z)^2)^{3/2}} \,.
\end{equation*}
 For BTZ black holes and AdS space, there is another relation which is given by
\begin{equation} \label{BTZrel}
z f'(z) -{2 f(z)+2}=0 \,.
\end{equation}
Using this relation, 
it can be seen that AdS space and BTZ black holes have the same value of the bulk part, and same structure of the boundary part.
Using the profile of the boundary $Q$,
$x'(z)$ obtained from (\ref{BTZTT}),
boundary parts for  AdS space and BTZ black holes have the same
value, $T$, i.e., (\ref{BTZav}). However, the results of the new type black holes are
definitely  different from these cases.

Now, in order to find the profile of $Q$ in new type black holes,  let us consider the
boundary condition  given by (\ref{B-condi}). Using the
relations (\ref{rel}),  we have two equations for $\tau \tau$ and
$zz$ components, which are separately given by
%
%
\begin{eqnarray}\label{tt}
0&=& \frac{z^2}{L^3 f(z) \left(1+f(z)\, x'(z)^2\right)^{3/2}} \Bigg[  \left(\sigma +\frac{1}{2  m^2  L^2}\right) f(z)^2 x'(z)^3 +  \sigma \left(f(z)-\frac{1}{2} z f'(z) \right)x'(z)  \nonumber \\
&& +\frac{1 }{2 m^2 L^2 } {\left(f(z) \left(7-6 f(z)+2 z f'(z)\right) + \frac{1}{2} z f'(z)  \left(z f'(z)+1\right) \right) x'(z) } \, , \nonumber \\
&&-z f(z) \left(\sigma -\frac{1-2 f(z)+z f'(z)}{2 m^2  L^2}\right)
x''(z) -L \,T \left(1+f(z) \,x'(z)^2\right)^{3/2} \Bigg] \,,
\\
%
0&=&
 \frac{z^2 f(z)}{2 L^3 \left(1+f(z) \,x'(z)^2\right)^{3/2}} \Bigg[ \sigma  \left(2 f(z)-z f'(z)\right) x'(z)- 2 L \,T \sqrt{1+f(z)  \, x'(z)^2}  \nonumber \\
&& + \frac{1}{ 2  m^2 L^2} \left(2 f(z)-z f'(z)-1\right) \left(2
f(z)-z f'(z)\right) x'(z) \Bigg] \,.
\end{eqnarray}
%
%
%
 %
Therefore from $z z$ component,
 we obtain
\begin{equation}   \label{nt}
T= \frac{ 2 f(z)-z f'(z) }{2 L \sqrt{1+f(z) \,x'(z)^2}}\left(\sigma
+\frac{2
   f(z)-z f'(z)-1}{2 m^2 L^2 }\right) x'(z) \,.
\end{equation}
This equation gives us the profile of $Q$. For BTZ black holes, with
Eq.(\ref{BTZrel}), it is easy to see that Eq.(\ref{nt}) reduces to
Eq.(\ref{BTZT}).
However, we should check if this is a correct solution of boundary
condition by checking $\tau \tau $ component. For BTZ black holes,
with $x'(z)$ obtained from (\ref{BTZTT}), it is easy to check that
Eq.(\ref{tt}) is satisfied.
%
For new type black holes at the special parameters, however,
 the $\tau \tau$ component  with   $x'(z)$ from Eq.(\ref{nt}) gives us the following condition. 
%
\begin{eqnarray} \label{ttcd}
\frac{1}{L^2 \left(z f'(z)-2 f(z)\right)^{6}}\Bigg[ {4 T z^2 \left(2
f(z)-z f'(z)-2\right) \left(\left(z f'(z)-2 f(z)\right)^4-6 \, L^2
\,T^2 f(z)\right)} \Bigg]=0 \,,
\end{eqnarray}
where we put $\sigma=1$ and $2 m^2 L^2=1$ for simplicity.
The first term of the numerator vanishes for BTZ black holes,
because of the relation (\ref{BTZrel}).
But new type black holes do not satisfy the relation. Therefore  for the new type black hole there is no solution to satisfy the boundary condition with constant $T$.
%
%

Next, let us consider the energy conservation  for the new type
black holes.
\begin{equation}
\nabla^j T^Q_{i j} ={1 \over {8 \pi G}} \Bigg[ { \Big( \sigma + {1 \over 2}
\hat s -{1 \over 2} \hat f \Big) (-N R^{\eta}_i) +(K \gamma_{ij}
-K_{i j} ) \nabla^j \Big( \sigma + {1 \over 2} \hat s -{1 \over 2}
\hat f  \Big) + \nabla^j H_{ij}   } \Bigg]\,,
\end{equation}
where we used the second Gauss-Codacci equation
(\ref{Gauss-Codazzi})
and  
\begin{eqnarray}
H^{ij} {\equiv}  -\nabla^{(i} \hat h ^{j)}+{1 \over 2} D_{\eta} \hat
f^{i j} + K_k^{(i} \hat f^{j) k} +\gamma^{i j} \Big( \nabla_k \hat
h^k -{1 \over 2} D_{\eta} \hat f \Big) \,.
\end{eqnarray}
Using Eq.(\ref{rel}), we find that
\begin{eqnarray}
\nabla^j T^Q_{\tau j} &=&0  \,,  \\
\label{ec} \nabla^j T^Q_{z j} &=&   {1 \over {8 \pi G}} \Bigg[ {
\Big( \sigma + {1 \over 2} \hat s -{1 \over 2} \hat f \Big) (-N
R^{\eta}_z) } { -\frac{{ \left(2 f(z)-z f'(z)-2\right) \left(2
f(z)-z f'(z)+4\right) x'(z)}}{4 m^2 L^3 z \left(1+f(z)
x'(z)^2\right)^{3/2}}    } \Bigg]
   \nonumber \\
 &=& \frac{\left(2 f(z)-z f'(z)-2\right) x'(z)
}{16 G L    \pi  z
   \left(1+f(z) \, x'(z)^2\right)^{3/2}}  \Bigg[ \sigma (1+  f(z)
  \, x'(z)^2 )    \nonumber \\
&&+{1 \over {2 m^2 L^2} } \biggl( { \left(2 f(z)-z f'(z)-1\right)
f(z)x'(z)^2 - 4 f(z)+2 z f'(z)-1} \biggr) \Bigg] \,,
\end{eqnarray}
where
\begin{eqnarray}
N= \frac{L}{z \sqrt{1+f(z) x'(z)^2}} \, \quad , \quad R^{\eta}_z =
\frac{\left(2-2 f(z)+z f'(z)\right) }{2 L^2} x'(z) \,.
\end{eqnarray}
For BTZ black hole case, it is easy to check that eq.(\ref{ec}) is
zero from eq.(\ref{BTZrel}).

%
%

%
%
%

\newpage


\begin{thebibliography}{99}


\bibitem{Maldacena:1997re}
  J.~M.~Maldacena,
  {\it The Large N limit of superconformal field theories and supergravity},
 {Adv. Theor. Math. Phys.} {\bf 2} (1998) 231
  ; { Int. J. Theor. Phys.}  {\bf 38} (1999) 1113
 [arXiv:9711200 [hep-th]].

\bibitem{Witten:1998qj}
  E.~Witten,
  {\it Anti-de Sitter space and holography},
 {Adv. Theor. Math. Phys.}  {\bf 2} (1998) 253
 [arXiv:9802150 [hep-th]].
;
  S.~S.~Gubser, I.~R.~Klebanov and A.~M.~Polyakov,
  {\it Gauge theory correlators from noncritical string theory},
{\PL}  B {\bf 428} (1998) 105
  [arXiv:hep-th/9802109].



\bibitem{'tHooft:1993gx}
  G.~'t Hooft,
  {\it Dimensional reduction in quantum gravity},
  [arXiv:gr-qc/9310026].


\bibitem{Susskind:1994vu}
  L.~Susskind,
  {\it The World As A Hologram},
  J.\ Math.\ Phys.\  {\bf 36} (1995) 6377
  [arXiv:hep-th/9409089].

\bibitem{Strominger:1997eq}
  A.~Strominger,
  {\it Black hole entropy from near-horizon microstates},
  JHEP {\bf 9802} (1998) 009
  [arXiv:hep-th/9712251].

\bibitem{Strominger:1996sh}
  A.~Strominger and C.~Vafa,
  {\it Microscopic Origin of the Bekenstein-Hawking Entropy},
  Phys.\ Lett.\  B {\bf 379} (1996) 99
  [arXiv:hep-th/9601029].


\bibitem{Takayanagi:2011zk}
  T.~Takayanagi,
  {\it Holographic Dual of BCFT},
  Phys.\ Rev.\ Lett.\  {\bf 107} (2011) 101602
  [arXiv:1105.5165 [hep-th]]
;
  M.~Fujita, T.~Takayanagi and E.~Tonni,
  {\it Aspects of AdS/BCFT},
  JHEP {\bf 1111} (2011) 043
  [arXiv:1108.5152 [hep-th]]
;
  M.~Alishahiha and R.~Fareghbal,
  {\it Boundary CFT from Holography},
  Phys.\ Rev.\ D {\bf 84} (2011) 106002
  [arXiv:1108.5607 [hep-th]].




\bibitem{Affleck:1991tk}
  I.~Affleck and A.~W.~W.~Ludwig,
  {\it Universal noninteger `ground state degeneracy' in critical quantum
  systems},
  Phys.\ Rev.\ Lett.\  {\bf 67} (1991) 161.

\bibitem{Friedan:2003yc}
  D.~Friedan and A.~Konechny,
  {\it On the boundary entropy of one-dimensional quantum systems at low
   temperature},
  Phys.\ Rev.\ Lett.\  {\bf 93} (2004) 030402
  [arXiv:hep-th/0312197].

\bibitem{Zamolodchikov:1986gt}
  A.~B.~Zamolodchikov,
  {\it Irreversibility of the Flux of the Renormalization Group in a 2D Field
  Theory},
  JETP Lett.\  {\bf 43} (1986) 730 ;
  Pisma Zh.\ Eksp.\ Teor.\ Fiz.\  {\bf 43} (1986) 565.



\bibitem{Deser:1982}
  R.~Jackiw, S.~Templeton and S.~Deser,
  {\it Three-Dimensional Massive Gauge Theories},
   \PRL {\bf48} (1982) 975.


\bibitem{Deser:1981wh}
  S.~Deser, R.~Jackiw and S.~Templeton,
  {\it Topologically massive gauge theories},
   {\it Ann. Phys.}  {\bf 140} (1982) 372
  [Erratum-ibid.\ 1988 {\bf 185} 406 , 1988 {\bf 281} 409].




\bibitem{Li:2008dq}
  W.~Li, W.~Song and A.~Strominger,
  {\it Chiral Gravity in Three Dimensions},
  JHEP {\bf 0804} (2008) 082
  [arXiv:0801.4566 [hep-th]].




\bibitem{Grumiller:2008qz}
  D.~Grumiller and N.~Johansson,
  {\it Instability in cosmological topologically massive gravity at the chiral
  point},
  JHEP {\bf 0807} (2008) 134
  [arXiv:0805.2610 [hep-th]].


\bibitem{Bergshoeff:2009hq}
  E.~A.~Bergshoeff , O.~Hohm and P.~K.~Townsend,
  {\it Massive Gravity in Three Dimensions},
  {\PRL} {\bf 102}  (2009) 201301
  [arXiv:0901.1766 [hep-th]].


\bibitem{Sinha:2010}
  A.~Sinha,
  {\it On the new massive gravity and AdS/CFT},
   {\JHEP}{\bf 1006} (2010) 061
  [arXiv:1003.0683 [hep-th]].


\bibitem{Gullu:2010}
  I.~Gullu, T.~C.~Sisman and B.~Tekin,
  {\it Born-Infeld extension of new massive gravity},
   \CQG {\bf 27} (2010) 162001
  [arXiv:1003.3935 [hep-th]].

\bibitem{Lu:2011zk}
  H.~Lu and C.~N.~Pope,
  {\it Critical Gravity in Four Dimensions},
  \PRL {\bf 106} (2011) 181302
  [arXiv:1101.1971 [hep-th]]
;
  H.~Lu, Y.~Pang and C.~N.~Pope,
  {\it Conformal Gravity and Extensions of Critical Gravity},
  Phys.\ Rev.\ D {\bf 84} (2011) 064001
  [arXiv:1106.4657 [hep-th]].




\bibitem{Deser:2011xc}
  S.~Deser, H.~Liu, H.~Lu, C.~N.~Pope, T.~C.~Sisman and B.~Tekin,
  {\it Critical Points of D-Dimensional Extended Gravities},
  \PR D {\bf 83}  (2011) 061502
  [arXiv:1101.4009 [hep-th]].



\bibitem{deBoer:2011wk}
  J.~de Boer, M.~Kulaxizi and A.~Parnachev,
  {\it Holographic Entanglement Entropy in Lovelock Gravities},
  JHEP {\bf 1107} (2011) 109
  [arXiv:1101.5781 [hep-th]].

\bibitem{Hung:2011xb}
  L.~Y.~Hung, R.~C.~Myers and M.~Smolkin,
  {\it On Holographic Entanglement Entropy and Higher Curvature Gravity},
  JHEP {\bf 1104} (2011) 025
  [arXiv:1101.5813 [hep-th]].

\bibitem{Ogawa:2011fw}
  N.~Ogawa and T.~Takayanagi,
  {\it Higher Derivative Corrections to Holographic Entanglement Entropy for AdS
  Solitons},
  JHEP {\bf 1110} (2011) 147
  [arXiv:1107.4363 [hep-th]].

\bibitem{Bergshoeff:2009aq}
   E.~A.~Bergshoeff , O.~Hohm and P.~K.~Townsend,
   {\it More on Massive 3D Gravity},
   {\PR} D {\bf 79}  (2009) 124042
  [arXiv:0905.1259 [hep-th]].

\bibitem{Townsend:2009}
  E.~A.~Bergshoeff , O.~Hohm and P.~K.~Townsend,
  {\it On Higher Derivatives in 3D Gravity and Higher Spin Gauge Theories},
  {Ann. Phys.} {\bf 325}  (2010) 1118
  [arXiv:0911.3061 [hep-th]].

\bibitem{Bergshoeff:2009fj}
  E.~A.~Bergshoeff , O.~Hohm and P.~K.~Townsend,
  {\it On massive gravitons in 2+1 dimensions},
  [arXiv:0912.2944 [hep-th]].

\bibitem{Kwon:2011jz}
  Y.~Kwon, S.~Nam, J.~D.~Park and S.~H.~Yi,
  {\it Holographic Renormalization and Stress Tensors in New Massive Gravity},
  JHEP {\bf 1111} (2011) 029
  [arXiv:1106.4609 [hep-th]].


\bibitem{Kwon:2011ey}
  Y.~Kwon, S.~Nam, J.~D.~Park and S.~H.~Yi,
  {\it Quasi Normal Modes for New Type Black Holes in New Massive Gravity},
  \CQG {\bf  28} (2011) 145006
  [arXiv:1102.0138 [hep-th]].


\bibitem{Nam:2010dd}
  S.~Nam, J.~D.~Park and S.~H.~Yi,
  {\it AdS Black Hole Solutions in the Extended New Massive Gravity},
   {\JHEP}{\bf 1007} (2010) 058
  [arXiv:1005.1619 [hep-th]].


\bibitem{Nam:2010ma}
  S.~Nam, J.~D.~Park  and S.~H.~Yi,
  {\it Mass and Angular momentum of Black Holes in New Massive Gravity},
   \PR D {\bf 82} (2010) 124049
  [arXiv:1009.1962 [hep-th]].


\bibitem{Cardy:2004hm}
  J.~L.~Cardy,
  {\it Boundary conformal field theory},
  [arXiv:hep-th/0411189].


\bibitem{Calabrese:2004eu}
   P.~Calabrese and J.~L.~Cardy,
   {\it Entanglement entropy and quantum field theory},
   J. Stat. Mech {\bf 0406} (2004) P06002
   [arXiv:hep-th/0405152].


\bibitem{Azeyanagi:2007qj}
  T.~Azeyanagi, A.~Karch, T.~Takayanagi and E.~G.~Thompson,
  {\it Holographic Calculation of Boundary Entropy},
  JHEP {\bf 0803} (2008) 054
  [arXiv:0712.1850 [hep-th]].

\bibitem{Brown:1992br}
  J.~D.~Brown and J.~W.~York,
  {\it Quasilocal energy and conserved charges derived from the gravitational
  action},
   \PR  D {\bf 47} (1993) 1407
  [arXiv:9209012 [hep-th]].


\bibitem{Balasubramanian:1999re}
  V.~Balasubramanian and P.~Kraus,
  {\it A Stress tensor for Anti-de Sitter gravity},
  {Commun.\ Math.\ Phys.}  {\bf 208} (1999) 413
  [arXiv:hep-th/9902121].



\bibitem{Myers:2010tj}
  R.~C.~Myers and A.~Sinha,
  {\it Holographic c-theorems in arbitrary dimensions},
    {\JHEP}{\bf 1101} (2011) 125
  [arXiv:1011.5819[hep-th]].


\bibitem{Hohm:2010jc}
  O.~Hohm and E.~Tonni,
  {\it A boundary stress tensor for higher-derivative gravity in AdS and Lifshitz
  backgrounds},
  {\JHEP} {\bf 1004} (2010) 093
  [arXiv:1001.3598 [hep-th]].




\bibitem{Banados:1992wn}
  M.~Ba\~{n}ados, C.~Teitelboim and J.~Zanelli,
  {\it The Black hole in three-dimensional space-time},
  \PRL {\bf 69} (1992) 1849
  [arXiv:hep-th/9204099].


\bibitem{Nojiri:1999mh}
  S.~Nojiri and S.~D.~Odintsov,
  {\it On the conformal anomaly from higher derivative gravity in AdS/CFT
  correspondence},
  Int.\ J.\ Mod.\ Phys.\  A {\bf 15} (2000) 413
  [arXiv:hep-th/9903033].




\bibitem{Liu:2009bk}
  Y.~Liu and Y.~W.~Sun,
  {\it Note on New Massive Gravity in $AdS_3$},
  JHEP {\bf 0904} (2009) 106
  [arXiv:0903.0536 [hep-th]].








\bibitem{Chiodaroli:2011fn}
  M.~Chiodaroli, E.~D'Hoker and M.~Gutperle,
  {\it Simple holographic duals to boundary CFTs},
  [arXiv:1111.6912 [hep-th]].






\bibitem{Myers:2010xs}
  R.~C.~Myers and A.~Sinha,
  {\it Seeing a c-theorem with holography},
  Phys.\ Rev.\  D {\bf 82} (2010) 046006
  [arXiv:1006.1263 [hep-th]].


\bibitem{Oliva:2009ip}
  J.~Oliva, D.~Tempo  and R.~Troncoso,
  {\it Three-dimensional black holes, gravitational solitons, kinks and wormholes
  for BHT masive gravity},
   {\JHEP}{\bf 0907} (2009) 011
  [arXiv:0905.1545 [hep-th]]
;
  A.~Perez, D.~Tempo and R.~Troncoso,
  {\it Gravitational solitons, hairy black holes and phase
  transitions in BHT massive gravity},
  {\JHEP}{\bf 1107} (2011) 093
  [arXiv:1106.4849 [hep-th]].




\bibitem{Maldacena:2011mk}
  J.~Maldacena,
  {\it Einstein Gravity from Conformal Gravity},
  [arXiv:1105.5632 [hep-th]].



\bibitem{Hyun:2011ej}
  S.~Hyun, W.~Jang, J.~Jeong and S.~H.~Yi,
  {\it Noncritical Einstein-Weyl Gravity and the AdS/CFT Correspondence},
  [arXiv:1111.1175 [hep-th]].




\bibitem{Metsaev}
R.~R.~Metsaev, {\it Stueckelberg approach to 6d conformal gravity},
Talk given at  workshop "Supersymmetries and Qauntum Symmetries",
July 18-23, Dubna, Russia ;  v3  of   {\it Ordinary-derivative
formulation of conformal low-spin fields}, [arXiv:0707.4437
[hep-th]].





\bibitem{Brown:1986nw}
  J.~D.~Brown and M.~Henneaux,
  {\it Central Charges in the Canonical Realization of Asymptotic Symmetries: An
  Example from Three-Dimensional Gravity},
 {Commun. Math. Phys.}  {\bf 104} (1986) 207.

















%


















































































\end{thebibliography}
\end{document}